\documentclass[12pt]{article}

\usepackage{a4wide}
\newcommand{\defeq}{\stackrel{\Delta}{=}}
\newcommand{\ket}[1]{| #1 \rangle}
\newcommand{\PAR}{\mbox{{\sf PAR}}}
\newcommand{\Tr}{\mbox{{\rm Tr} }}
\newcommand{\ketbra}[1]{| #1 \rangle \langle #1 |}
\newcommand{\trace}[1]{\| #1 \|_t}
\newcommand{\ellone}[1]{\| #1 \|_1}

\newtheorem{definition}{Definition} 
\newtheorem{proposition}{Proposition} 
\newtheorem{theorem}{Theorem}    
\newtheorem{lemma}{Lemma}      
\newcommand{\qed}{\hfill{$\rule{6pt}{6pt}$}} 
\newenvironment{proof}{\noindent{\bf Proof}:}{\qed}

\title{{\bf Lower bounds in the quantum cell probe model}\thanks{
A preliminary version of this paper appeared in the 
{\em Proceedings of the 28th International Colloquium
on Automata, Languages and Programming}, Lecture Notes in
Computer Science, vol. 2076, pages 358--369, 2001.}
}

\author{
Pranab~Sen\thanks{
Laboratoire de Recherche en Informatique, 
Universit\'{e} de Paris-Sud, 91405 Orsay, France. 
Email: {\sf pranab@lri.fr}.
Most of this work was done while visiting UC Berkeley and 
DIMACS, under a Sarojini Damodaran 
International Fellowship grant, while the author was a graduate 
student at the
Tata Institute of Fundamental Research, Mumbai, India.} 
\and 
S.~Venkatesh\thanks{
Center for Discrete Mathematics and Theoretical Computer Science, 
Rutgers University, Piscataway, NJ 08854, USA.
Email: {\sf venkat@dimacs.rutgers.edu}.
Most of this work was done while the author was a post doctoral
researcher at the Institute of Advanced Study, Princeton, USA,
supported by NSF grant CCR-9987845 and a joint IAS-DIMACS post doctoral
fellowship.}
}

\date{}

\begin{document}

\maketitle

\begin{abstract}
We introduce a new model for studying quantum data structure problems
--- the {\em quantum cell probe model}. We prove a lower bound
for the {\em static predecessor} problem in the {\em address-only}
version of this model where we allow quantum
parallelism only over the `address lines' of the queries. The
address-only quantum cell probe model
subsumes the classical cell probe model, and many quantum query
algorithms like Grover's algorithm fall into this framework. 
Our lower bound improves the previous known lower bound for the
predecessor problem in the classical cell probe model with 
randomised query schemes, and matches the classical deterministic 
upper bound of Beame and Fich~\cite{beame:predecessor}. Beame
and Fich~\cite{beame:predecessor} have also
proved a matching lower bound for the predecessor problem, but only 
in the classical deterministic setting. Our
lower bound has the advantage that it holds for the more general
quantum model, and also, its proof is substantially simpler than
that of Beame and Fich.

We prove our lower bound by obtaining a round elimination lemma for 
quantum communication complexity. A similar lemma was proved by 
Miltersen, Nisan, Safra and Wigderson~\cite{miltersen:roundelim} 
for classical communication complexity, but it was not strong
enough to prove a lower bound matching the upper bound of Beame
and Fich. Our quantum round elimination lemma 
also allows us to
prove rounds versus communication tradeoffs for some quantum
communication complexity problems like the `greater-than' problem.

We also study the {\em static membership} problem in the quantum
cell probe model. Generalising a result of Yao~\cite{yao:tablesort}, we
show that if the storage scheme is {\em implicit}, that is it
can only store members of the subset and `pointers', then any quantum
query scheme must make $\Omega(\log n)$ probes. 
\end{abstract}

\section{Introduction}
\label{sec:intro}
A static data structure problem
consists of a set of data $D$, a set of queries $Q$, a set of
answers $A$, and a function
$f:D \times Q \rightarrow A$. The aim is to store the data efficiently
and succinctly,
so that any query can be answered with only a few probes to the data
structure.
In a seminal paper~\cite{yao:tablesort}, Yao introduced 
the (classical) {\em cell probe
model} for studying static data structure problems (in the
classical setting). Thereafter, this model
has been used extensively to prove (classical)
upper and lower bounds for several
data structure problems (see e.g.~\cite{fredman:hashing,
miltersen:roundelim, beame:predecessor, buhrman:bitprobe}). 
A classical {\em $(s,w,t)$ cell probe scheme} for $f$
has two components: a {\em storage scheme} and a {\em query
scheme}. Given the data $d \in D$ to be stored,
the storage scheme stores it as a table $T_d$
of $s$ cells, each cell $w$ bits
long. $w$ is called the word size of the scheme.
The query scheme has to answer queries.
Given a query $q \in Q$, the query scheme
computes the answer $f(d,q)$
to that query by making at most $t$ probes to the
stored table $T_d$, where each probe reads one cell at a time. 
The storage
scheme is deterministic whereas the query scheme can be deterministic
or randomised.  The goal is to study tradeoffs between $s$, $t$ and
$w$. For an overview of results in this model, see the survey by
Miltersen~\cite{miltersen:cellprobe}.

In this paper, we study static data structure problems, such as the
{\em static membership} problem and the 
{\em static predecessor} problem, when the query
algorithm is allowed to query the table using a quantum
superposition. We formalise this by defining the {\em quantum cell
probe model} similar to the {\em quantum bit probe model}
of Radhakrishnan, Sen and Venkatesh~\cite{radhakrishnan:quantset}. 
Informally, in the quantum cell probe model, the storage scheme is
classical deterministic as before and stores the data $d \in D$
as a table of cells $T_d$; however, the query scheme is quantum
and can query the table $T_d$ using a quantum superposition.
We show a lower bound for the
predecessor problem in a restricted version of this model, which
we call the {\em address-only} quantum cell probe model. In the
predecessor problem, the storage scheme has to store a subset $S$ of
size at most $n$ from the universe $[m]$, such that given any query
element $x \in [m]$, one can quickly find the predecessor of $x$ in $S$.
\begin{quote}
{\bf Result 1 (Lower bound for predecessor, informal statement)}
{\it
Suppose we have an address-only quantum
cell probe solution with constant probability of error
for the static predecessor problem, where the 
universe size is $m$ and the subset size is at most $n$, using 
$n^{O(1)}$ cells of storage with word size $(\log m)^{O(1)}$  
bits. Then the number of
queries is at least 
$\Omega \left( \frac{\log \log m}{\log \log \log m} \right)$ 
as a function of $m$, and at least 
$\Omega \left( \sqrt{\frac{\log n}{\log \log n}} \right)$ 
as a function of $n$.
\/}
\end{quote}

We then consider the static membership problem. Here one has to
answer membership queries instead of predecessor queries.
Yao~\cite{yao:tablesort}
showed that if the universe is large enough, any classical 
cell probe solution with an {\em
implicit} deterministic storage  scheme and a deterministic query
scheme for the static membership problem must 
make $\Omega(\log n)$ probes to the table in the worst case.
An {\em implicit} storage scheme either stores
a `pointer value' (viz. a value which
is not an element of the universe) or an element of $S$ in a cell.
In particular, it is not allowed to store an element of the universe
which is not a member of $S$.
We generalise Yao's result to the quantum
setting.
\begin{quote}
{\bf Result 2 (Lower bound for membership, informal statement)}
{\it 
Suppose we have a quantum cell probe solution with an implicit
storage scheme for the static membership problem. Then,
if the universe is large
enough compared to the number of cells of storage, the size of the
universe of `pointers' and the size
of the stored subset, the 
query algorithm must make $\Omega(\log n)$ probes, even if we allow
constant probability of error.
\/}
\end{quote}

\paragraph{Remarks:} \ \\
1.\ \ Our {\em address-only} quantum cell probe model
subsumes the classical cell probe model with randomised query
schemes. Hence, our lower
bound for the static predecessor problem
also holds in this setting. This improves the previous lower bound
$\Omega(\sqrt{\log \log m})$ as a function of $m$
and $\Omega(\log^{1/3} n)$ as a function of $n$ for this setting,
shown by Miltersen, Nisan, Safra and 
Wigderson~\cite{miltersen:roundelim}. 
Beame and Fich~\cite{beame:predecessor} have shown an upper bound
matching our lower bound up to constant factors, which uses
$n^{O(1)}$ cells of storage of word size $O(\log m)$ bits. 
In fact, both the storage and the query schemes are
classical deterministic in Beame and Fich's solution. 
In their paper,
Beame and Fich~\cite{beame:predecessor} also show a lower bound of
$t = \Omega \left( \frac{\log \log m}{\log \log \log m} \right)$ as
a function of $m$ for
$(n^{O(1)}, 2^{(\log m)^{1 - \Omega(1)}}, t)$ classical
deterministic cell probe schemes,
and a lower bound of 
$t = \Omega \left( \sqrt{\frac{\log n}{\log \log n}} \right)$ as
a function of $n$ for
$(n^{O(1)}, (\log m)^{O(1)}, t)$ classical
deterministic cell probe schemes.
But their lower bound proof breaks down
if the query scheme is randomised.
Our result thus shows that the upper bound
scheme of Beame and Fich is optimal all
the way up to the bounded error address-only quantum cell 
probe model. Also, our proof is 
substantially simpler than that of Beame and Fich. \\
2.\ \ It is known that querying in superposition gives a speed up over
classical algorithms for certain data retrieval problems, the most
notable one being Grover's algorithm~\cite{grover:search} for
searching an unordered list of $n$ elements using $O(\sqrt{n})$
quantum queries. The power of quantum querying for data structure
problems was studied in the context of static membership 
by Radhakrishnan, Sen and
Venkatesh~\cite{radhakrishnan:quantset}. In their paper, they worked in
the quantum bit probe model, which is our quantum cell probe model
where the word size is just one bit.  They showed, roughly speaking,
that quantum querying does not give much advantage over classical
schemes for the set membership problem. Our result above seems to
suggest that quantum search is perhaps not more powerful than
classical search for the predecessor problem as well. \\
3.\ \ In the next section, we formally describe the 
``address-only'' restrictions we
impose on the query algorithm. Informally, they amount to this: {\em
we allow quantum parallelism over the `address lines' going into the
table, but we have a fixed quantum state on the `data lines'.}  This
restriction on quantum querying does not make the model trivial. In
fact, many non-trivial quantum search algorithms, such as Grover's
algorithm~\cite{grover:search}, Farhi {\em et al.}'s 
algorithm~\cite{farhi:binsearch}, and H{\o}yer {\em et al.}'s 
algorithm~\cite{hoyer:binsearch}, already satisfy these restrictions. \\
4.\ \ For the static membership problem, Fredman, Koml\'{o}s and
Szemer\'{e}di~\cite{fredman:hashing} have shown a 
classical deterministic cell probe solution where the storage scheme
uses $O(n)$ cells of word size $O(\log m)$ bits, and the query
scheme makes only a constant number of probes. In this solution,
the storage scheme may store elements of the universe in the table
which are not members of the subset to be stored. Hence the
restriction that the storage scheme be implicit is necessary for any
such result. We note that implicit storage 
schemes include
many of the standard storage schemes like sorted array, hash table,
search trees etc.

\subsection{Techniques}
\label{subsec:techniques}
The lower bounds for the static membership problem shown in the quantum
bit probe model by 
Radhakrishnan {\em et al.}~\cite{radhakrishnan:quantset} 
relied on linear algebraic
techniques. Unfortunately, these techniques appear to be powerless in
the quantum cell probe model. In fact, to show the lower bound above
for the static predecessor problem,
we use a connection between quantum 
data structure problems and two-party quantum
communication complexity, similar to what was used by
Miltersen, Nisan, Safra and Wigderson~\cite{miltersen:roundelim},
and Beame and Fich~\cite{beame:predecessor} 
for showing their (classical) lower
bounds. Miltersen {\em et al.}~\cite{miltersen:roundelim}
proved a technical lemma in classical
communication complexity called
the {\em round elimination lemma} and
derived from it lower bounds for various static data structure
problems, including the predecessor problem. But their round elimination
lemma was not strong enough to prove a lower bound matching the
upper bound of Beame and Fich.
In this paper we prove a stronger (!) round
elimination lemma for the quantum communication complexity model,
which we then use to show a quantum
lower bound for the static predecessor problem matching Beame and
Fich's upper bound. Our quantum round elimination lemma is proved
using quantum information theoretic techniques. Inspired by 
these techniques, we prove a still stronger round elimination lemma
in classical communication complexity.

We now give an informal description of the round elimination lemma.
Suppose $f:E \times F \rightarrow G$ is a function. In the
communication game corresponding to $f$, Alice gets a string 
$x \in E$, Bob gets a string $y \in F$, and they have to 
communicate and compute $f(x,y)$.  In
the communication game $f^{(n)}$, Alice gets $n$
strings $x_1, \ldots, x_n \in E$; Bob gets an integer $i \in [n]$, a
string $y \in F$, and a copy of the strings $x_1, \ldots,
x_{i-1}$. Their aim is to communicate and 
compute $f(x_i, y)$. Suppose a quantum protocol for
$f^{(n)}$ is given where Alice starts, and her first message 
is much smaller than $n$ qubits.  Intuitively, it would
seem that since Alice does not know $i$, the first round of
communication cannot give much information about $x_i$, and thus,
would not be very useful to Bob. Hence it should be possible to
eliminate the first round of 
communication, giving a quantum protocol for computing $f(x_i,y)$ where
Bob starts, with one less round of communication, and
having the same message complexity
and similar error probability. The round elimination lemma
justifies this intuition. Moreover, we show that this is true even if
Bob also gets copies of $x_1, \ldots, x_{i-1}$, a case which is needed
in many applications. 
\begin{quote}
{\bf Result 3 (Round elimination lemma, informal statement)}
{\it 
A $t$ round quantum protocol for
$f^{(n)}$ with Alice starting, where the first message of Alice
is much smaller than $n$ qubits, gives us a $t-1$ round 
quantum protocol for $f$
where Bob starts, with the same message complexity and similar error
probability. An analogous statement holds for classical randomised
protocols.
\/}
\end{quote}

Round reduction arguments have been given earlier
in quantum communication complexity, most notably
by Klauck, Nayak, Ta-Shma and Zuckerman~\cite{klauck:pointer}.
However, for technical reasons, the previous arguments do not go
far enough to prove lower bounds for the communication games arising
from data structure problems like the predecessor problem. We
need a technical quantum version of the round elimination lemma of
Miltersen {\em et al.}~\cite{miltersen:roundelim}, to prove the desired
quantum lower bounds.

The round elimination lemma also has applications
to other communication complexity problems,
which might be interesting on their own. 
For example, it can be used to prove rounds versus
communication tradeoffs for the `greater-than' problem.
In the `greater-than' problem $GT_n$, Alice is given $x \in \{0,1\}^n$,
Bob is given $y \in \{0,1\}^n$, and they have to communicate
and decide whether $x > y$ (treating $x,y$ as integers).
\begin{quote}
{\bf Result 4}
{\it 
The $t$ round bounded error quantum (classical randomised)
communication complexity of
$GT_n$ is $\Omega(n^{1/t} t^{-3})$ ($\Omega(n^{1/t} t^{-2})$).
\/}
\end{quote}
There exists a bounded error classical randomised protocol for
$GT_n$ using $t$ rounds of communication and having a complexity of
$O(n^{1/t} \log n)$. Hence, for a constant number of rounds, our
quantum lower bound matches the classical upper bound to within
logarithmic factors. For one round quantum protocols, our result implies
an $\Omega(n)$ lower bound for $GT_n$ (which is optimal to within
constant factors), improving upon the 
previous $\Omega(n/\log n)$ lower bound of 
Klauck~\cite{klauck:ccsurvey}. No rounds versus communication tradeoff
for this problem, for more than one round,
was known earlier in the quantum setting.
For classical randomised protocols, Miltersen 
{\em et al.}~\cite{miltersen:roundelim} showed a lower bound
of $\Omega(n^{1/t} 2^{-O(t)})$ using their round elimination lemma.
If the number of rounds is unbounded, then there is a classical
randomised protocol for $GT_n$  using $O(\log n)$ rounds of 
communication and having a complexity of 
$O(\log n)$~\cite{nisan:ccthreshold}. An
$\Omega(\log n)$ lower bound for the bounded error quantum communication
complexity of $GT_n$ (irrespective of the number of rounds)
follows from Kremer's
result~\cite{kremer:quantcc} that the bounded error quantum
communication complexity of a function is lower bounded (up to
constant factors) by the
logarithm of the one round (classical)
deterministic communication complexity.
%

\subsection{Organisation of the paper}
\label{subsec:organisation}
Section~\ref{sec:def} contains definitions of various terms that will
be used throughout the paper. In Section~\ref{sec:prelim}, we
discuss some lemmas that will be needed in the proofs of the main
theorems. Section~\ref{sec:roundelim} 
contains a proof of the quantum and classical round elimination lemmas.
Proofs of some lemmas
required to prove the round elimination lemma
proper have been relegated to the appendix.
In Section~\ref{sec:applications}, we apply our 
round elimination lemma to prove lower bounds 
for the query complexity of the static predecessor problem and
the communication complexity of the `greater-than' problem.
Section~\ref{sec:sorttable} contains a proof of our
lower bound for implicit storage quantum cell probe schemes for the
static membership problem. 
We conclude with a few remarks and 
some open problems in Section~\ref{sec:conclusion}.

\section{Definitions}
\label{sec:def}
In this section we define some of the terms which we will be
using in this paper. 

\subsection{The quantum cell probe model}
\label{subsec:quantcellprobe}

A quantum {\em $(s,w,t)$ cell probe scheme} for a static data structure 
problem $f: D \times Q \rightarrow A$
has two components: a classical deterministic {\em storage scheme} that 
stores the data $d \in D$ in a table $T_d$ using
$s$ cells each containing $w$ bits, and a quantum {\em query scheme}
that answers queries by `quantumly probing a cell at a time' $t$ times.
Formally speaking, the table $T_d$ is made available to the
query algorithm in the form of an oracle unitary transform $O_d$. To
define $O_d$ formally, we represent the basis states of the query
algorithm as
$\ket{j,b,z}$, where $j \in [s-1]$ is a binary string of
length $\log s$, $b$ is a binary string of length $w$, and 
$z$ is a binary string of
some fixed length. Here, 
$j$ denotes the address of a cell in the table $T_d$,
$b$ denotes the qubits which will hold the contents of a cell
and $z$ stands for the rest of the qubits (`work qubits')
in the query algorithm.
$O_d$ maps $\ket{j,b,z}$ to $\ket{j,b\oplus (T_d)_j,z}$, where
$(T_d)_j$ is a bit string of length $w$ and denotes the contents of the
$j$th cell in $T_d$.
A quantum query scheme with $t$ probes is just a sequence of
unitary transformations
\begin{displaymath}
U_{0} \rightarrow O_d \rightarrow U_{1} \rightarrow O_d
      \rightarrow \ldots U_{t-1}
      \rightarrow O_d \rightarrow U_{t}
\end{displaymath}
where $U_{j}$'s are arbitrary unitary transformations that do not
depend on $d$ (representing the internal computations
of the query algorithm). For a query $q \in Q$, the
computation starts in a computational basis state
$\ket{q}\ket{0}$, where we
assume that the ancilla qubits are initially in the basis state
$\ket{0}$. Then we apply in succession,
the operators $U_0, O_d, U_1, \ldots, U_{t-1}, 
O_d, U_t$, and measure the final state.
The answer consists of the values on some of the
output wires of the circuit. 
We say that the scheme has worst case error probability less than
$\epsilon$ if 
the answer is equal to $f(d,q)$, for
every $(d,q) \in D \times Q$, with
probability greater than $1 - \epsilon$. 
The term `bounded error quantum scheme' means that $\epsilon = 1/3$.

We now formally define the {\em address-only quantum cell probe model}.
Here the storage scheme is as in the general model,
but the query scheme is restricted to be `address-only'.
This means that the state vector before a query to the oracle $O_d$
is always a {\em tensor product} of a state vector 
on the address and work qubits (the $(j,z)$ part in 
$(j,b,z)$ above), and a state 
vector on the data qubits (the $b$ part in
$(j,b,z)$ above). The state vector on the data qubits before a
query to the oracle $O_d$ is {\em independent of the 
query element $q$ and the data $d$} but
can vary with the probe number. Intuitively, we are only
making use of quantum parallelism over the address lines of
a query. This
mode of querying a table subsumes classical querying, and also many 
non-trivial
quantum algorithms like Grover's algorithm~\cite{grover:search},
Farhi {\em et al.}'s algorithm~\cite{farhi:binsearch}, 
H{\o}yer {\em et al.}'s algorithm~\cite{hoyer:binsearch}
etc. satisfy the `address-only' condition.
For classical querying, the state vector on the data qubits is 
$\ket{0}$, independent of the probe number. 
For Grover and Farhi {\em et al.}, the state vector on the data qubit is
$(\ket{0} - \ket{1}) / \sqrt{2}$, independent of the probe number.
For H{\o}yer {\em et al.}, the state vector on the data qubit is 
$\ket{0}$ for some probe numbers, and 
$(\ket{0} - \ket{1}) / \sqrt{2}$ for the other probe numbers.

\subsection{Quantum communication protocols}
\label{subsec:defcc}
We consider two party quantum communication
protocols as defined by Yao~\cite{yao:quantcc}. 
Let $E,F,G$ be arbitrary finite sets and
$f: E \times F \rightarrow G$ be a function.
There are two players Alice and Bob, who hold qubits.
When the communication game starts, Alice holds
$\ket{x}$ where $x \in E$ together with some ancilla qubits in the
state $\ket{0}$, and Bob holds $\ket{y}$ where $y \in F$ together
with some ancilla qubits in the state $\ket{0}$. Thus the qubits
of Alice and Bob are initially in computational basis states, and
the initial superposition is simply
$\ket{x}_A \ket{0}_A \ket{y}_B \ket{0}_B$. Here the subscripts denote
the ownership of the qubits by Alice and Bob. The players take turns
to communicate to compute $f(x,y)$. Suppose it is Alice's turn.
Alice can make an
arbitrary unitary transformation on her qubits and then send one
or more qubits to Bob. Sending qubits does not change the overall
superposition, but rather changes the ownership of the qubits, allowing
Bob to apply his next unitary transformation on his original qubits
plus the newly received qubits. At the end of the protocol, the
last recipient of qubits performs a measurement on the qubits in
her possession to output an answer.
We say a quantum protocol computes $f$ with $\epsilon$-error in the
worst case,
if for any input $(x,y) \in E \times F$, the
probability that the protocol outputs
the correct result $f(x,y)$ is greater than $1 - \epsilon$.
The term `bounded error quantum protocol' means that $\epsilon = 1/3$.

We require that Alice and Bob
make a secure copy of their inputs before beginning the protocol.
This is possible since the inputs to Alice and Bob are in 
computational basis states.
Thus, without loss of generality,
the input qubits of Alice and Bob are never sent
as messages, their state remains unchanged throughout the protocol,
and they are never
measured i.e. some work qubits are measured to determine the result of
the protocol. 
We call such protocols {\em secure}. 
We will assume
henceforth that all our protocols are secure. 

We now define the concept of a {\em safe} 
quantum protocol, which will be used in the statement of the 
quantum round elimination lemma.
\begin{definition}[Safe quantum protocol] 
A $[t,c,l_1,\ldots,l_t]^A$ ($[t,c,l_1,\ldots,l_t]^B$) safe quantum 
protocol is a secure quantum protocol where Alice (Bob) starts
the communication, the first message is $l_1 + c$ qubits long,
the  $i$th message, for $i \ge 2$, is $l_i$ qubits long, and
the communication goes on for $t$ rounds. We think of the
first message as having two parts: the `main part' which is
$l_1$ qubits long, and the `safe overhead part' which is $c$ qubits
long. The density matrix of the `safe overhead' is independent of the 
inputs to Alice and Bob. 
\end{definition}
Later on in the paper, we also use the notation $(t,c,a,b)^A$ 
($(t,c,a,b)^B$) to
denote a $[t,c,l_1,\ldots,l_t]^A$ ($[t,c,l_1,\ldots,l_t]^B$)
safe quantum protocol, where the per round message lengths
of Alice and Bob are $a$ and $b$ qubits respectively i.e. if
Alice (Bob) starts, $l_i = a$ for $i$ odd and $l_i = b$ for
$i$ even ($l_i = b$ for $i$ odd and $l_i = a$ for $i$ even).

\paragraph{Remark:} 
The concept of a safe quantum protocol may look strange at first.
The reason we need to define it, intuitively speaking, is as follows.
The communication games arising from data structure problems
often have an asymmetry between the message lengths of Alice and Bob.
This asymmetry is crucial to prove lower bounds on the number
of rounds of communication.
In the previous quantum round reduction arguments (e.g. those
of Klauck {\em et al.}~\cite{klauck:pointer}), the complexity
of the first message in the protocol increases quickly as the
number of rounds is reduced and the asymmetry gets lost.
This leads to a problem where the
first message soon gets big enough to potentially
convey substantial information about the input of one player to the
other, destroying any hope of proving strong lower bounds on the
number of rounds.
The concept of a safe protocol allows us to get around this problem.
We show through a careful
quantum information theoretic analysis of the round reduction process,
that in a safe protocol,
though the complexity of the first message increases a lot, this
increase is confined to the safe overhead and so, the
information content does not increase much. This is the key property
which allows us to prove a round elimination lemma for safe
quantum protocols.

In this paper we will deal with quantum protocols with {\em public
coins}. Intuitively, a public coin quantum protocol is 
a probability distribution over finitely many ({\em coinless})
quantum protocols. We shall
henceforth call the standard definition of a quantum protocol
as {\em coinless}. Our definition
is similar to the classical scenario, where a randomised protocol
with public coins is a probability distribution over finitely
many deterministic
protocols. We note however, that our definition of a public coin quantum
protocol is {\em not} the same as that of a quantum protocol with prior
entanglement, which has been studied previously 
(see e.g.~\cite{cleve:ip}). Our definition is weaker, in that it
does not allow the unitary transformations of Alice and Bob to
alter the `public coin'.
\begin{definition}[Public coin quantum protocol]
In a quantum protocol with a public coin, there is, before the
start of the protocol, a 
quantum state called a {\em public coin}, of the form
$\sum_c \sqrt{p_c} \ket{c}_A \ket{c}_B$, where the
subscripts denote ownership of qubits by Alice and Bob, 
$p_c$ are finitely many non-negative real
numbers and $\sum_c p_c = 1$. 
Alice and Bob make (entangled) copies of their respective halves of the
public coin using {\sc CNOT} gates before commencing the
protocol. The unitary transformations of Alice and Bob during the
protocol do not touch the public coin.
The public coin is never measured, nor is it ever sent as a message.
\end{definition}
Hence, one can
think of the public coin quantum protocol to be a probability
distribution, with probability $p_c$, over finitely many
coinless quantum protocols
indexed by the coin basis states $\ket{c}$. A {\em safe public coin}
quantum protocol is similarly defined as
a probability distribution over finitely many safe
coinless quantum protocols.

\paragraph{Remarks:} \  \\ 
1.\ \ We need to define public coin quantum protocols in order to 
make use of the harder direction of Yao's minimax 
lemma~\cite{yao:minimax}. The minimax lemma
is the main tool which allows us to convert `average case'
round reduction arguments to `worst case' arguments. We need `worst
case' round reduction arguments in proving lower 
bounds for the rounds
complexity of communication games arising from data
structure problems. This is because many of these lower
bound proofs
use some notion of ``self-reducibility'', arising from the 
original data
structure problem, which fails to hold in the `average case' but holds
for the `worst case'. The quantum round reduction arguments of 
Klauck {\em et al.}~\cite{klauck:pointer}
are `average case' arguments, and this is 
one of the reasons why they do not suffice to prove lower bounds
for the rounds complexity of communication games arising from data
structure problems. \\
2.\ \  Parallel repetitions of protocols, as well as constructing
new protocols from old ones using both the directions of 
Yao's minimax lemma, preserve the ``safety'' property.

For an input $(x,y) \in E \times F$, we define the error 
$\epsilon^{P}_{x,y}$ of 
the protocol $P$ on $(x,y)$, to be the probability
that the result of $P$ on input $(x,y)$ is not equal to $f(x,y)$. For
a protocol $P$, given a probability distribution
$D$ on $E \times F$, we define the 
average error 
$\epsilon^{P}_D$ of $P$ with respect to $D$ as the expectation
over $D$ of the error of $P$ on inputs $(x,y) \in E \times F$. 
We define 
$\epsilon^P$ to be worst case error of $P$ on inputs 
$(x,y) \in E \times F$.

\section{Preliminaries}
\label{sec:prelim}
In this section we state some facts which will be useful in what
follows.

\subsection{Yao's minimax lemma}
For completeness, we state Yao's minimax lemma~\cite{yao:minimax}
for safe
quantum protocols in the (slightly more general) flavour that will be 
required by us. The proof of this flavour of the lemma is very similar
to the standard proof, using the von Neumann minimax theorem.
\begin{lemma}[Yao's minimax lemma]
\label{lem:minimax}
Consider $[t,c,l_1,\ldots,l_t]^A$ safe quantum protocols $P$ 
for a function $f: E \times F \rightarrow G$. Let
$D$ denote a probability distribution on the inputs 
$(x,y) \in E \times F$. Then
\begin{displaymath}
\inf_{\mbox{$P$: {\it public coin\/}}} \epsilon^P =
\sup_D \inf_{\mbox{$P$: {\it coinless\/}}} \epsilon^P_D =
\sup_D \inf_{\mbox{$P$: {\it public coin\/}}} \epsilon^P_D
\end{displaymath}
Analogous properties hold for classical protocols too.
\end{lemma}

\subsection{Quantum cell probe complexity and communication}
\label{subsec:cellcomm}
In this subsection, we describe the connection between the quantum
cell probe complexity of a static data structure problem and 
the quantum communication
complexity of an associated communication game. 
Let $f : D \times Q \rightarrow A$ be a static data structure problem.
Consider a two-party
communication problem where Alice is given a query $q \in Q$,
Bob is given data $d \in D$, and they have to communicate and
find out the answer $f(d,q)$. We have the following lemma, which
is a quantum analogue of a lemma of Miltersen~\cite{miltersen:union}
relating cell probe complexity to communication complexity in
the classical setting.
\begin{lemma}
\label{lem:queryandcc}
Suppose we have a quantum $(s,w,t)$ cell 
probe solution to the static data structure problem $f$.
Then we have a
$(2t, 0, \log s + w, \log s + w)^A$ safe coinless quantum
protocol for the corresponding
communication problem. If the query scheme is address-only,
we can get a $(2t, 0, \log s, \log s + w)^A$ safe coinless 
quantum protocol.
The error probability of the communication protocol is the same as
that of the cell probe scheme.
\end{lemma}
\begin{proof} 
Given a quantum $(s,w,t)$ cell
probe solution to the static data structure problem $f$,
we can get a 
$(2t, 0, \log s + w, \log s + w)^A$ safe coinless quantum
protocol for the corresponding
communication problem by just
simulating the cell probe solution. If in addition,
the query scheme is address-only, the messages from Alice to
Bob need consist only of the `address' part. This can be
seen as follows. Let the state vector of the data qubits before
the $i$th query be $\ket{\theta_i}$. $\ket{\theta_i}$ is independent
of the query element and the stored data. 
Bob keeps $t$ special ancilla registers in states 
$\ket{\theta_i}, 1 \leq i \leq t$ at the start of the protocol $P$.
These special ancilla registers are in tensor with the rest of the
qubits of Alice and Bob at the start of $P$.
Protocol $P$ simulates the cell probe solution, but with the following
modification. To simulate the $i$th query of the cell probe
solution, Alice prepares her `address' and `data' qubits as in the
query scheme, but sends the `address' qubits only. Bob treats those
`address' qubits together with $\ket{\theta_i}$ in the $i$th
special ancilla register as Alice's query, and performs the oracle
table transformation on them. He then sends these qubits (both
the `address' as well as the $i$th special register qubits) to Alice.
Alice exchanges the contents of the $i$th special register with
her `data' qubits (i.e. exchanges the basis states), and proceeds
with the simulation of the query scheme. This gives us a 
$(2t, 0, \log s, \log s + w)^A$ safe coinless 
quantum protocol with the same 
error probability as
that of the cell probe query scheme.
\end{proof}

In many natural data structure problems $\log s$ is much smaller than
$w$ and thus, in the address-only quantum case, we get a 
$(2t, 0, \log s, O(w))^A$ safe protocol. In the classical
setting, one gets a $(2t, 0, \log s, w)^A$ protocol. 
This asymmetry in message lengths is crucial
in proving non-trivial lower bounds on $t$.
The concept of a safe quantum protocol helps us in exploiting
this asymmetry.

\subsection{Background from quantum information theory}
\label{subsec:quantinfo}
In this subsection, we discuss some basic facts 
from quantum information theory that will be used
in the proof of the round elimination lemma. We follow the notation
of Klauck, Nayak, Ta-Shma and Zuckerman's paper~\cite{klauck:pointer}.
For a good account of quantum information theory, see the
book by Nielsen and Chuang~\cite{nielsen:quant}.

If $A$ is a quantum system with density matrix $\rho$, then
$S(A) \defeq S(\rho) \defeq -\Tr \rho \log \rho$ is the 
{\em von Neumann entropy} of $A$.
If $A,B$ are
two disjoint quantum systems, their {\em mutual information} 
is defined as $I(A:B) \defeq S(A) + S(B) - S(AB)$.
We now state some properties about von Neumann entropy and
mutual information which will be
useful later. The proofs follow easily from the definitions,
using basic properties of 
von Neumann entropy like subadditivity and triangle inequality
(see e.g.~\cite[Chapter~11]{nielsen:quant}).

\begin{lemma}
\label{lem:mutualinfo}
Suppose $A,B,C$ are disjoint quantum systems. Then
\begin{displaymath}
I(A:BC) = I(A:B) + I(AB:C) - I(B:C)
\end{displaymath}
\begin{displaymath}
0 \leq I(A:B)  \leq 2 S(A) 
\end{displaymath}
If the Hilbert space of $A$ has dimension $d$, then
\begin{displaymath}
0 \leq S(A) \leq \log d
\end{displaymath}
\end{lemma}

Suppose $X,Q$ are disjoint quantum systems with finite dimensional
Hilbert spaces ${\cal H}, {\cal K}$ respectively. 
For every computational basis state $\ket{x} \in {\cal H}$, suppose 
$\sigma_x$ is a density matrix in ${\cal K}$.
Suppose the density matrix of $(X,Q)$ is
$\sum_x p_x \ketbra{x} \otimes \sigma_x$, where $p_x > 0$ and
$\sum_x p_x = 1$. Thus 
$X$ is in a mixed state $\{p_x, \ket{x} \}$, and we
shall say that $X$ is a classical random variable and that
$Q$ is a quantum encoding $\ket{x} \mapsto \sigma_{x}$ of $X$.
Define $\sigma \defeq \sum_x p_x \sigma_x$. 
$\sigma$ is the reduced
density matrix of $Q$, and we shall say that $\sigma$ is the
the density matrix of the average encoding. 
Then, $S(XQ) = S(X) + \sum_x p_x S(\sigma_x)$, and hence,
$I(X:Q) = S(\sigma) - \sum_x p_x S(\sigma_x)$.

Let $X,Y,Q$ be disjoint quantum systems with finite dimensional
Hilbert spaces ${\cal H}, {\cal K}, {\cal L}$ respectively.
Let $x \in {\cal H}$, $y \in {\cal K}$ be computational basis
vectors. For every $\ket{x}\ket{y} \in {\cal H} \otimes {\cal K}$, 
suppose $\sigma_{xy}$ is a density matrix in ${\cal L}$.
Let $Z$ refer to the quantum system $(X,Y)$.
Suppose $(X,Y,Z)$ has density matrix
$\sum_{x,y} p_{xy} \ketbra{x} \otimes \ketbra{y} \otimes \sigma_{xy}$, 
where $p_{xy} > 0$ and $\sum_{x,y} p_{xy} = 1$.
Thus, $X$ and $Y$ are classical random variables, and
$Z = XY$ is in a mixed state
$\{p_{xy}, \ket{x}\ket{y} \}$.
$Q$ is a quantum encoding 
$\ket{xy} \mapsto \sigma_{xy}$ of $Z$. Define 
$q^x_y$ to be the (conditional) probability that $Y=y$ given that
$X=x$. $\ket{y} \mapsto \sigma_{xy}$
can be thought of as a quantum encoding $Q^x$ of $Y$ given that $X=x$.
The joint density matrix of $(Y, Q^x)$ is
$\sum_y q^x_y \ketbra{y} \otimes \sigma_{xy}$.
We let $I((Y:Q)|X = x)$ denote the mutual information of this
encoding.

We now prove the following propositions.

\begin{proposition}
\label{prop:safe}
Let $M_1,M_2$ be disjoint finite dimensional quantum systems.
Suppose $M \defeq (M_1, M_2)$ is a quantum encoding 
$\ket{x} \mapsto \sigma_x$ of a 
classical random variable $X$. Suppose the density matrix of $M_2$ is
independent of $X$ i.e. $\mbox{Tr}_{M_1}\, \sigma_x$ is the same
for all $x$. Let $M_1$ be supported on $a$ qubits.
Then, $I(X:M) \leq 2a$. 
\end{proposition}
\begin{proof}
By Lemma~\ref{lem:mutualinfo}, 
$I(X:M) = I(X:M_1 M_2) = I(X:M_2) + I(X M_2:M_1) - I(M_2:M_1)$.
But since the density matrix of $M_2$ is independent of 
$X$, $I(X:M_2)=0$. Hence, by again using Lemma~\ref{lem:mutualinfo},
we get that $I(X:M) \leq I(X M_2:M_1) \leq 2 S(M_1) \leq 2a$.
\end{proof} 

\paragraph{Remarks:} \ \\
1. \ \ This proposition is the key observation allowing us to
``ignore'' the size of the ``safe'' overhead $M_2$ 
in the round elimination
lemma. It will be very useful in the applications of the round
elimination lemma, where the complexity of the first message
in the protocol increases quickly, but the blow up is confined to
the ``safe'' overhead. Earlier round reduction arguments were unable
to handle this large blow up in the complexity of the first message.\\
2. \ \ In the above proposition, if $M$ is a 
classical encoding of $X$ (i.e. an encoding 
$\ket{x} \mapsto \sigma_x$, where $\sigma_x$ is a density matrix
of a mixture of computational basis vectors), we get the improved
inequality $I(X:M) \leq a$.

The next proposition has been observed by Klauck
{\em et al.}~\cite{klauck:pointer}.
\begin{proposition}
\label{prop:additivity}
Suppose $M$ is a quantum encoding of a classical
random variable $X$. Suppose $X = X_1 X_2 \ldots X_n$, where the
$X_i$ are classical independent random variables. Then,
$I(X_1 \ldots X_n:M) = \sum_{i=1}^n I(X_i:M X_1 \ldots X_{i-1})$. 
\end{proposition}
\begin{proof}
The proof is by induction on $n$, using Lemma~\ref{lem:mutualinfo}
repeatedly. We also use the fact that
$I(X_i:X_{i+1} \ldots X_n)=0$ for $1 \leq i < n$,
since $X_1, \ldots, X_n$ are independent classical random variables.
\end{proof} 

\begin{proposition}
\label{prop:averaging}
Let $X,Y$ be classical random variables and
$M$ be a quantum encoding of $(X,Y)$. Then,
$I(Y:M X) = I(X:Y) + E_{X} [I((Y:M)|X=x)]$. 
\end{proposition}
\begin{proof}
Let $\sigma_{xy}$ be the density matrix of $M$ when $X,Y=x,y$.
Let $p_x$ be the (marginal)
probability that $X=x$ and $q^x_y$ the (conditional)
probability that $Y=y$ given $X=x$. 
Define $\sigma_x \defeq \sum_y q^x_y \sigma_{xy}$. We now have
\begin{eqnarray*}
I(Y:M X) & = & S(Y) + S(M X) - S(M X Y) \\
         & = & S(Y) + S(X) + \sum_x p_x S(\sigma_x) -
	       (S(X Y) + \sum_{x,y} p_x q^x_y S(\sigma_{xy})) \\
	 & = & I(X:Y) +
	       \sum_x p_x (S(\sigma_x) - \sum_y q^x_y S(\sigma_{xy})) \\
	 & = & I(X:Y) + \sum_x p_x I((Y:M)|X=x) \\
	 & = & I(X:Y) + E_{X} [I((Y:M)| X=x)]
\end{eqnarray*}
\end{proof}

\section{The round elimination lemmas}
\label{sec:roundelim}

In this section we prove our round elimination lemmas for safe
public coin quantum protocols and public coin classical randomised
protocols. Since a public coin quantum protocol can
be converted to a coinless quantum protocol at the expense of an
additional ``safe'' overhead in the first message, we also get
a similar round elimination lemma for coinless protocols. We 
can decrease the overhead to logarithmic in the total bit size
of the inputs by a technique similar to the public to private coins
conversion for classical randomised protocols~\cite{newman:private}.
But since the statement of the round elimination lemma is cleanest
for safe public coin quantum protocols, we give it below 
for such protocols only. Similar remarks apply to the classical
setting.

\subsection{The quantum round elimination lemma}
\label{subsec:quantroundelim}

In this subsection we prove our round elimination lemma for
safe public coin quantum protocols.
We first state the following round reduction
lemma, which can be proved in a
manner similar to the proof of Lemma~4.4 in Klauck 
{\em et al.}~\cite{klauck:pointer}, but with a careful accounting of 
``safe'' overheads in the messages communicated by Alice and Bob.
Intuitively speaking, the lemma says that if the first message of Alice
carries little information about her input, under some probability
distribution on inputs, then it can be eliminated, giving rise
to a protocol where Bob starts, with one less 
round of communication, and the same message complexity and similar
error probability, with respect to the same probability
distribution on inputs.
We observe, in the lemma below,
that though there is a overhead of
$l_1 + c$ qubits on the first message of Bob, it is a ``safe'' overhead.
\begin{lemma}
\label{lem:quantroundreduce}
Suppose $f: E \times F \rightarrow G$ is a function. Let $D$ be
a probability distribution on $E \times F$, and
$P$ be a $[t,c,l_1,\ldots,l_t]^A$ safe coinless 
quantum protocol for $f$. 
Let $X$ stand for the classical random variable denoting Alice's input
(under distribution $D$),
$M$ be the first message of Alice in the protocol $P$, and 
$I(X:M)$ denote the mutual information between $X$ and $M$ under
distribution $D$. Then
there exists a $[t-1,c+l_1,l_2,\ldots,l_t]^B$ safe coinless 
quantum protocol $Q$ for $f$, such that
\begin{displaymath}
\epsilon^Q_D \leq \epsilon^P_D + ((2 \ln 2) I(X:M))^{1/4}
\end{displaymath}
\end{lemma}
A proof of this lemma can be found in the appendix.

We can now prove the quantum round elimination lemma 
(for the communication game $f^{(n)}$).
\begin{lemma}[Quantum round elimination lemma] 
\label{lem:quantroundelim}
Suppose $f:E \times F \rightarrow G$ is a function.
Suppose the communication game $f^{(n)}$ has a 
$[t,c,l_1,\ldots,l_t]^A$ safe public coin quantum protocol
with worst case error less than $\delta$.
Then there is a $[t-1,c+l_1,l_2,\ldots,l_t]^B$ safe public coin
quantum protocol for $f$ with worst case error less than 
$\epsilon \defeq \delta + (4 l_1 \ln 2 /n)^{1/4}$.
\end{lemma}
\begin{proof}
Suppose the given protocol for $f^{(n)}$ has worst case error
$\tilde{\delta} < \delta$. Define 
$\tilde{\epsilon} \defeq \tilde{\delta} + (4 l_1 \ln 2 /n)^{1/4}$.
To prove the quantum round elimination lemma
it suffices to give,
by the harder direction of the minimax lemma (Lemma~\ref{lem:minimax}), 
for any probability distribution $D$ on $E \times F$, a 
$[t-1,c+l_1,l_2,\ldots,l_t]^B$ safe public coin quantum 
protocol $P$ for $f$ with average distributional error 
$\epsilon^P_D \leq \tilde{\epsilon} < \epsilon$. 
To this end, we will first construct a probability distribution 
$D^\ast$ on $E^n \times [n] \times F$ as follows.
Choose $i \in [n]$ uniformly at random. Choose independently, for 
each $j \in [n]$,
$(x_j,y_j) \in E \times F$ according to distribution $D$. Set
$y=y_i$ and throw away $y_j, j \neq i$.
By the easier direction of the minimax
lemma (Lemma~\ref{lem:minimax}), we get a 
$[t,c,l_1,\ldots,l_t]^A$ safe coinless quantum protocol $P^\ast$
for $f^{(n)}$ with distributional error, 
$\epsilon^{P^\ast}_{D^\ast} \leq \tilde{\delta} < \delta$. 
In $P^\ast$, Alice gets $x_1,\ldots,x_n$, Bob gets
$i$, $y$ and $x_1,\ldots,x_{i-1}$.
We shall construct the desired protocol $P$
from the protocol $P^\ast$.

Let $M$ be the first message of Alice in $P^\ast$. By the definition
of a safe protocol, $M$ has two parts: $M_1$ $l_1$ qubits long, and
the ``safe'' overhead $M_2$, $c$ qubits long. 
Let the input to Alice be denoted by
the classical random variable $X=X_1 X_2 \ldots X_n$ where 
$X_i$ is the classical
random variable corresponding to the $i$th input to Alice. 
Let the classical random variable $Y$ denote the input 
$y$ of Bob.
Define $\epsilon^{P^\ast}_{D^\ast;i;x_1,\ldots,x_{i-1}}$
to be the average error of $P^\ast$ under distribution $D^\ast$
when $i$ is fixed and $X_1,\ldots,X_{i-1}$ are fixed
to $x_1,\ldots,x_{i-1}$. 
Using Propositions~\ref{prop:safe}, \ref{prop:additivity}, 
\ref{prop:averaging} and 
the fact that under distribution
$D^\ast$, $X_1,\ldots,X_n$ are
independent classical random variables, we get that 
\begin{equation}
\label{eq:info1}
\begin{array}{ccl}
\frac{2 l_1}{n} 
     & \geq & \frac{I (X:M)}{n} \\
     &   =  & E_i [I (X_i:M X_1, \ldots, X_{i-1})] \\
     &   =  & E_{i,X} [I ((X_i:M)|X_1,\ldots,
                                          X_{i-1}=x_1,\ldots,x_{i-1})]
\end{array}
\end{equation}
Also
\begin{equation}
\label{eq:error1}
\tilde{\delta} \geq \epsilon^{P^\ast}_{D^\ast} =
       E_{i,X} \left[\epsilon^{P^\ast}_{D^\ast;i;x_1,\ldots,x_{i-1}}
	       \right]
\end{equation}
The expectations above are under distribution $D^\ast$. 

For any $i \in [n]$, $x_1,\ldots,x_{i-1} \in E$, define
the $[t,c,l_1,\ldots,l_t]^A$
safe coinless quantum protocol $P'_{i;x_1,\ldots,x_{i-1}}$
for the function $f$ as follows. Alice is given
$x \in E$ and Bob is given $y \in F$. Bob sets
$i$ to the given value, and both Alice and Bob set
$X_1, \ldots, X_{i-1}$ to
the values $x_1, \ldots, x_{i-1}$. Alice puts an
independent copy of a pure state $\ket{\psi}$ (defined below)
for each of the
inputs  $X_{i+1},\ldots,X_n$. She sets $X_i=x$ and Bob sets 
$Y=y$. Then they run protocol $P^\ast$ on these inputs. Here 
$\ket{\psi} \defeq \sum_{x \in E} \sqrt{p_x} \ket{x}$, where 
$p_x$ is the (marginal) probability of $x$ under distribution $D$.
Since $P^\ast$ is a safe coinless quantum protocol, so is 
$P'_{i;x_1,\ldots,x_{i-1}}$. Because $P^\ast$ is a secure protocol,
the probability that $P'_{i;x_1,\ldots,x_{i-1}}$
makes an error for an input $(x,y)$, 
$\epsilon^{P'_{i;x_1,\ldots,x_{i-1}}}_{x,y}$,
is the average probability of
error of $P^\ast$ under distribution $D^\ast$ when 
$i$ is fixed to the given value, $X_1,\ldots,X_{i-1}$ are fixed
to $x_1,\ldots,x_{i-1}$, and $X_i,Y$ are fixed to $x,y$.
Hence, the average probability of error of $P'_{i;x_1,\ldots,x_{i-1}}$ 
under distribution $D$
\begin{equation}
\label{eq:error2}
\epsilon^{P'_{i;x_1,\ldots,x_{i-1}}}_D = 
       \epsilon^{P^\ast}_{D^\ast;i;x_1,\ldots,x_{i-1}}
\end{equation}
Let $M'$ denote the first message of $P'_{i;x_1,\ldots,x_{i-1}}$ and
$X'$ denote the register $X_i$ holding the input $x$ to Alice. 
Because of the ``secureness'' of $P^\ast$, the density matrix
of $(X',M')$ in protocol $P'_{i;x_1,\ldots,x_{i-1}}$ is the
same as the density matrix of $(X_i,M)$ in protocol $P^\ast$ when
$X_1,\ldots,X_{i-1}$ are set to $x_1,\ldots,x_{i-1}$. Hence
\begin{equation}
\label{eq:info2}
I (X':M') = 
      I ((X_i:M)|X_1,\ldots, X_{i-1}=x_1,\ldots,x_{i-1})
\end{equation}

Using Lemma~\ref{lem:quantroundreduce} and equations (\ref{eq:error2})
and (\ref{eq:info2}), we get a 
$[t-1,c+l_1,l_2,\ldots,l_t]^B$ safe 
coinless quantum protocol $P_{i;x_1,\ldots,x_{i-1}}$ for $f$ with
\begin{equation}
\label{eq:average1}
\begin{array}{cll}
\epsilon^{P_{i;x_1,\ldots,x_{i-1}}}_D 
       & \leq & \!\!\!\!\epsilon^{P'_{i;x_1,\ldots,x_{i-1}}}_D +
                    ((2 \ln 2) I(X':M'))^{1/4} \\
       &   =  & \!\!\!\!\epsilon^{P^\ast}_{D^\ast;
                                           i;x_1,\ldots,x_{i-1}} + 
                    ((2 \ln 2) I ((X_i:M)|
                       X_1,\ldots, X_{i-1}=x_1,\ldots,x_{i-1}))^{1/4} 
\end{array}
\end{equation}

We now construct a $[t-1,c+l_1,l_2,\ldots,l_t]^B$ safe
public coin quantum protocol $P$ for $f$, which is nothing but
a probability distribution (under $D^\ast$) over the 
safe coinless quantum protocols $P_{i;x_1,\ldots,x_{i-1}}$,
$i \in [n]$, $x_1,\ldots,x_{i-1} \in E$. For protocol $P$, we get
(note that the expectations below are under distribution $D^\ast$)
\begin{eqnarray*}
\epsilon^P_D 
     &   =  & E_{i,X} \left[\epsilon^{P_{i;x_1,\ldots,x_{i-1}}}_D 
                      \right] \\
     & \leq & E_{i,X} \left[\epsilon^{P^\ast}_{D^\ast;i;
                                        x_1,\ldots,x_{i-1}} +
                            \left((2 \ln 2) I ((X_i:M)|
                                       X_1,\ldots, X_{i-1}=
                                          x_1,\ldots,x_{i-1})
                            \right)^{1/4}
                      \right]                              \\
     & \leq & E_{i,X} \left[\epsilon^{P^\ast}_{D^\ast;i;
                                        x_1,\ldots,x_{i-1}}
                      \right] +
              \left((2 \ln 2) 
                    E_{i,X} \left[I ((X_i:M)|
                                       X_1,\ldots, X_{i-1}=
                                            x_1,\ldots,x_{i-1})
                            \right]
              \right)^{1/4}                                \\
     & \leq & \tilde{\delta} +
                 \left(\frac{4 l_1 \ln 2}{n} \right)^{1/4} \\
     &   =  & \tilde{\epsilon}
\end{eqnarray*}
The first inequality follows from (\ref{eq:average1}), the second
inequality follows from the concavity of the fourth root function
and the last inequality from from (\ref{eq:info1}) and
(\ref{eq:error1}).

This completes the proof of the quantum round elimination lemma.
\end{proof}

\subsection{The classical round elimination lemma}
\label{subsec:classicalroundelim}

The proof of the classical round elimination lemma is similar to
that of the quantum round elimination lemma. First, we have the
following classical analogue of Lemma~\ref{lem:quantroundreduce}.
\begin{lemma}
\label{lem:classicalroundreduce}
Suppose $f: E \times F \rightarrow G$ is a function. Let $D$ be
a probability distribution on $E \times F$, and
$P$ be a $[t,0,l_1,\ldots,l_t]^A$ private coin
classical randomised protocol for $f$. 
Let $X$ stand for the classical random variable denoting Alice's input
(under distribution $D$),
$M$ be the first message of Alice in the protocol $P$, and 
$I(X:M)$ denote the mutual information between $X$ and $M$ under
distribution $D$. Then
there exists a $[t-1,0,l_2,\ldots,l_t]^B$ public coin
classical randomised protocol $Q$ for $f$, such that
\begin{displaymath}
\epsilon^Q_D \leq \epsilon^P_D + \frac{1}{2} ((2 \ln 2) I(X:M))^{1/2}
\end{displaymath}
\end{lemma}
A proof of the lemma is given in the appendix.

We can now prove the classical round elimination lemma
(for the communication game $f^{(n)}$).
\begin{lemma}[Classical round elimination lemma] 
\label{lem:classicalroundelim}
Suppose $f:E \times F \rightarrow G$ is a function.
Suppose the communication game $f^{(n)}$ has a 
$[t,0,l_1,\ldots,l_t]^A$ public coin classical randomised protocol
with worst case error less than $\delta$.
Then there is a $[t-1,0,l_2,\ldots,l_t]^B$ public coin
classical randomised protocol for $f$ with worst case error less than 
$\epsilon \defeq \delta + (1/2) (2 l_1 \ln 2 /n)^{1/2}$.
\end{lemma}
\begin{proof} {\bf (Sketch)}
The proof is similar to that of Lemma~\ref{lem:quantroundelim},
but using Lemma~\ref{lem:classicalroundreduce} instead of 
Lemma~\ref{lem:quantroundreduce}. 
Suppose the given protocol for $f^{(n)}$ has worst case error
$\tilde{\delta} < \delta$. Define 
$\tilde{\epsilon} \defeq \tilde{\delta} + (1/2) (2 l_1 \ln 2 /n)^{1/2}$.
To prove the classical round elimination lemma
it suffices to give,
by the harder direction of the minimax lemma (Lemma~\ref{lem:minimax}), 
for any probability distribution $D$ on $E \times F$, a 
$[t-1,0,l_2,\ldots,l_t]^B$ public coin classical randomised 
protocol $P$ for $f$ with average distributional error 
$\epsilon^P_D \leq \tilde{\epsilon} < \epsilon$. 
To this end, we construct the probability distribution 
$D^\ast$ on $E^n \times [n] \times F$ as before. By the easier 
direction of the minimax lemma (Lemma~\ref{lem:minimax}), we get a 
$[t,0,l_1,\ldots,l_t]^A$ classical deterministic protocol $P^\ast$
for $f^{(n)}$ with distributional error,
$\epsilon^{P^\ast}_{D^\ast} \leq \tilde{\delta} < \delta$. 
In $P^\ast$, Alice gets $x_1,\ldots,x_n \in E$, Bob gets
$i \in [n]$, $y \in F$ and a copy of $x_1,\ldots,x_{i-1}$.
We shall construct the desired protocol $P$
from the protocol $P^\ast$.

Let $M$ be the first message of Alice in $P^\ast$. 
Let the input to Alice be denoted by
the classical random variable $X=X_1 X_2 \ldots X_n$ where 
$X_i$ is the classical
random variable corresponding to the $i$th input to Alice. 
Let the classical random variable $Y$ denote the input 
$y$ of Bob.
Define $\epsilon^{P^\ast}_{D^\ast;i;x_1,\ldots,x_{i-1}}$
to be the average error of $P^\ast$ under distribution $D^\ast$
when $i$ is fixed and $X_1,\ldots,X_{i-1}$ are fixed
to $x_1,\ldots,x_{i-1}$. Arguing as before, we get
\begin{displaymath}
\frac{l_1}{n} 
      =  E_{i,X} [I ((X_i:M)|X_1,\ldots,
                                      X_{i-1}=x_1,\ldots,x_{i-1})]
\end{displaymath}
Also
\begin{displaymath}
\tilde{\delta} \geq \epsilon^{P^\ast}_{D^\ast} =
       E_{i,X} \left[\epsilon^{P^\ast}_{D^\ast;i;x_1,\ldots,x_{i-1}}
	       \right]
\end{displaymath}
The expectations above are under distribution $D^\ast$. 

For any $i \in [n]$, $x_1,\ldots,x_{i-1} \in E$, define
the $[t,0,l_1,\ldots,l_t]^A$
private coin classical randomised protocol $P'_{i;x_1,\ldots,x_{i-1}}$
for the function $f$ as follows. Alice is given
$x \in E$ and Bob is given $y \in F$. Bob sets
$i$ to the given value, and both Alice and Bob set
$X_1, \ldots, X_{i-1}$ to
the values $x_1, \ldots, x_{i-1}$. Alice tosses her private coin
to choose $X_{i+1},\ldots,X_n \in E$, where each
$X_j, i+1 \leq j \leq n$ is chosen independently according to the
(marginal) distribution on $E$ induced by $D$.
Alice sets $X_i=x$ and Bob sets 
$Y=y$. Then they run protocol $P^\ast$ on these inputs. 
The probability that $P'_{i;x_1,\ldots,x_{i-1}}$
makes an error for an input $(x,y)$, 
$\epsilon^{P'_{i;x_1,\ldots,x_{i-1}}}_{x,y}$,
is the average probability of
error of $P^\ast$ under distribution $D^\ast$ when 
$i$ is fixed to the given value, $X_1,\ldots,X_{i-1}$ are fixed
to $x_1,\ldots,x_{i-1}$, and $X_i,Y$ are fixed to $x,y$.
Hence, the average probability of error of $P'_{i;x_1,\ldots,x_{i-1}}$ 
under distribution $D$
\begin{displaymath}
\epsilon^{P'_{i;x_1,\ldots,x_{i-1}}}_D = 
       \epsilon^{P^\ast}_{D^\ast;i;x_1,\ldots,x_{i-1}}
\end{displaymath}
Let $M'$ denote the first message of $P'_{i;x_1,\ldots,x_{i-1}}$ and
$X'$ denote the register $X_i$ holding the input $x$ to Alice. Then
\begin{displaymath}
I (X':M') = 
      I ((X_i:M)|X_1,\ldots, X_{i-1}=x_1,\ldots,x_{i-1})
\end{displaymath}

Using Lemma~\ref{lem:classicalroundreduce} and arguing as before,
we can complete the proof of the classical round elimination
lemma.
\end{proof}

\section{Applications of the round elimination lemma}
\label{sec:applications}
In this section, we apply our round
elimination lemmas to prove lower bounds for the query
complexity of the static predecessor problem, and rounds versus 
communication tradeoffs for the `greater-than' problem.

\subsection{Static predecessor problem}
The proof of our lower bound for the static predecessor
problem in the address-only quantum cell probe model is
similar to the classical proof in Miltersen {\em et 
al.}~\cite{miltersen:roundelim}.
But because we use a stronger
round elimination lemma, we can prove stronger lower bounds.
We start by some preliminary observations.

\begin{definition}[Rank parity communication games, 
                   \cite{miltersen:roundelim}]
In the {\em rank parity} communication
game $\PAR_{p,q}$,
Alice is given a bit string $x$ of
length $p$, Bob is given a set $S$ of bit strings of length $p$,
$|S| \leq q$, and they have to communicate and decide
whether the rank of $x$ in $S$ (treating the bit 
strings as integers) is odd or even. 
By the rank of $x$ in $S$, we mean the cardinality of the
set $\{y \in S \mid y \leq x \}$.
In the game $\PAR^{(k),A}_{p,q}$,
Alice is given $k$ bit strings $x_1,\ldots,x_k$ each of 
length $p$, Bob is given a set $S$ of bit strings of length $p$,
$|S| \leq q$, an index $i \in [k]$, and a copy of $x_1,\ldots,x_{i-1}$;
they have to communicate and decide 
whether the rank of $x_i$ in $S$ is odd or even.
In the game $\PAR^{(k),B}_{p,q}$,
Alice is given a bit string $x$ of 
length $p$ and an index $i \in [k]$, Bob is given $k$ sets
$S_1,\ldots,S_k$ of bit strings of length $p$,
$|S_j| \leq q$;
they have to communicate and decide 
whether the rank of $x$ in $S_i$ is odd or even.
\end{definition}

\begin{proposition}
\label{prop:pred2rank}
Let there be a $(n^{O(1)},(\log m)^{O(1)},t)$ address-only
quantum cell probe solution 
to the static predecessor problem, where the universe size
is $m$ and the subset size is at most $n$.
Then there is a 
$\left(2t+O(1), 0, O(\log n), (\log m)^{O(1)} \right)^A$ 
safe coinless (and hence, public coin) quantum 
protocol for the rank parity communication game $\PAR_{\log m, n}$.
The error probability of the communication protocol is the same as
that of the cell probe scheme.
\end{proposition}
\begin{proof}
Consider the {\em static rank parity} data structure problem where 
the storage scheme has to store a set $S \subseteq [m]$, $|S| \leq n$,
and the query scheme, given a query $x \in [m]$, has to decide whether
the rank of $x$ in $S$ is odd or even.
Fredman, Koml\'{o}s and Szemer\'{e}di~\cite{fredman:hashing}
have shown the existence of two-level 
perfect hash tables containing, for each member $y$ of the
stored subset $S$, $y$'s rank in $S$, and
using $O(n)$ cells of word size $O(\log m)$ and requiring
only $O(1)$ classical deterministic cell probes.
Combining a $(n^{O(1)}, (\log m)^{O(1)}, t)$ address-only quantum
cell probe solution to the static predecessor problem with such
a perfect hash table, gives us a 
$(n^{O(1)} + O(n), \max((\log m)^{O(1)}, O(\log m)), t + O(1))$ 
address-only quantum
cell probe solution to the static rank parity problem.
The error probability of the cell probe scheme for the rank
parity problem is the same as the error probability of the cell
probe scheme for the predecessor problem.
By Lemma~\ref{lem:queryandcc}, we get a
$(2t+O(1), 0, O(\log n), (\log m)^{O(1)})^A$ safe coinless 
quantum protocol for the rank
parity communication game $\PAR_{\log m, n}$. 
The error probability of the communication protocol is the same as
that of the cell probe scheme for the predecessor problem.
\end{proof}

\begin{proposition}[\cite{miltersen:roundelim}]
\label{prop:rankred1}
Suppose $k$ divides $p$.
A communication protocol for $\PAR_{p,q}$ with Alice starting, 
gives us a communication protocol for $\PAR^{(k),A}_{p/k,q}$ with 
Alice starting, with the same message complexity, number of rounds 
and error probability.
\end{proposition}
\begin{proof}
Consider the problem $\PAR^{(k),A}_{p/k,q}$.
Alice, who is given $x_1,\ldots,x_k$, computes the concatenation
$\hat{x} \defeq x_1 \cdot x_2 \cdots x_k$. Bob, who is given
$S$, $i$ and $x_1,\ldots,x_{i-1}$, computes 
\begin{displaymath}
\hat{S} \defeq 
        \left\{ x_1 \cdot x_2 \cdots x_{i-1} 
                  \cdot y \cdot 0^{p(1 - i/k)} \mid  y \in S
        \right\}
\end{displaymath}
Alice and Bob then run the protocol for $\PAR_{p,q}$ on the inputs
$\hat{x}$, $\hat{S}$ to solve the problem $\PAR^{(k),A}_{p/k,q}$.
\end{proof}

\begin{proposition}[\cite{miltersen:roundelim}]
\label{prop:rankred2}
Suppose $k$ divides $q$, and $k$ is a power of $2$.
A communication protocol for $\PAR_{p,q}$ with Bob starting, 
gives us a communication protocol for 
$\PAR^{(k),B}_{p - \log k - 1,q/k}$ with 
Bob starting, with the same message complexity, number of rounds 
and error probability.
\end{proposition}
\begin{proof}
Consider the problem $\PAR^{(k),B}_{p - \log k - 1,q/k}$.
Alice, given $x$ and $i$, 
computes $\hat{x} \defeq (i-1) \cdot 0 \cdot x$.
Bob, given $S_1,\ldots,S_k$, computes the sets 
$S'_1,\ldots,S'_k$ where
\begin{displaymath}
S'_j \defeq  
        \left\{ \begin{array}{ll}
                  \left\{(j-1) \cdot 0 \cdot y \mid y \in S_j \right\} &
                            \mbox{if $|S_j|$ is even} \\
                  \left\{(j-1) \cdot 0 \cdot y \mid y \in S_j \right\} 
                   \bigcup \left\{(j-1) \cdot 1^{p - \log k} \right\} &
                            \mbox{if $|S_j|$ is odd}
                \end{array}
        \right.
\end{displaymath}
Above, the integers $(i-1), (j-1)$ are to be thought of as 
bit strings of length $\log k$.
Bob also computes $\hat{S} \defeq \bigcup_{j=1}^k S'_j$.
Alice and Bob then run the protocol for $\PAR_{p,q}$ on inputs
$\hat{x}$, $\hat{S}$ to solve the problem 
$\PAR^{(k),B}_{p - \log k - 1,q/k}$.
\end{proof}

We now prove the lower bound on the query complexity of static
predecessor in the address-only quantum cell probe model.
\begin{theorem}
\label{thm:quantpredlb}
Suppose we have a $(n^{O(1)}, (\log m)^{O(1)}, t)$ bounded error
quantum address-only
cell probe solution to the static predecessor problem, where the 
universe size is $m$ and the subset size is at most
$n$. Then the number of queries $t$ is at least 
$\Omega \left( \frac{\log \log m}{\log \log \log m} \right)$ 
as a function of $m$, and at least 
$\Omega \left( \sqrt{\frac{\log n}{\log \log n}} \right)$ 
as a function of $n$.
\end{theorem}
\begin{proof}
We basically imitate the proof of Miltersen 
{\em et al.}~\cite{miltersen:roundelim}, but in our quantum setting.
By Proposition~\ref{prop:pred2rank}, it suffices 
to consider communication protocols for the
rank parity communication game $\PAR_{\log m, n}$.
Let $n = 2^{(\log \log m)^2 / \log \log \log m}$.
Let $c_1 \defeq (4 \ln 2) 12^4$.
For any given constants $c_2, c_3 \geq 1$, define
\begin{displaymath}
a \defeq c_2 \log n ~~~~~
b \defeq (\log m)^{c_3} ~~~~~
t \defeq \frac{\log \log m}{(c_1 + c_2 + c_3) \log \log \log m}
\end{displaymath}
We shall show that the rank
parity communication game $\PAR_{\log m, n}$
does not have bounded error $(2t,0,a,b)^A$ 
safe public coin quantum protocols, thus proving the desired
lower bounds on the query complexity of static rank parity (and
hence, static predecessor) by Lemma~\ref{lem:queryandcc}. 

Given a $(2t,0,a,b)^A$ safe public coin quantum protocol for 
$\PAR_{\log m, n}$ with error probability $\delta$ ($\delta < 1/3$),
we get a $(2t,0,a,b)^A$ safe public coin quantum
protocol for 
\begin{displaymath}
\PAR^{(c_1 a t^4),A}_{\frac{\log m}{c_1 a t^4}, n}
\end{displaymath}
with the same error probability $\delta$,
by Proposition~\ref{prop:rankred1}. Using the quantum round elimination
lemma (Lemma~\ref{lem:quantroundelim}), we get a 
$(2t-1,a,a,b)^B$ safe public coin quantum protocol for
\begin{displaymath}
\PAR_{\frac{\log m}{c_1 a t^4}, n}
\end{displaymath}
but the error probability increases to at most $\delta + (12t)^{-1}$.
Using the reduction of Proposition~\ref{prop:rankred2}, we get
a $(2t-1,a,a,b)^B$ safe public coin quantum protocol for
\begin{displaymath}
\PAR^{(c_1 b t^4),B}_{\frac{\log m}{c_1 a t^4} - 
                     \log (c_1 b t^4) - 1, \frac{n}{c_1 b t^4}}
\end{displaymath}
with error probability at most $\delta + (12t)^{-1}$.
From the given values of the parameters, we see that
\begin{displaymath}
\frac{\log m}{(2 c_1 a t^4)^t} \geq \log (c_1 b t^4) + 1
\end{displaymath}
This implies that we also have a $(2t-1,a,a,b)^B$ safe public 
coin quantum protocol for
\begin{displaymath}
\PAR^{(c_1 b t^4),B}_{\frac{\log m}{2 c_1 a t^4}, \frac{n}{c_1 b t^4}}
\end{displaymath}
with error probability at most $\delta + (12t)^{-1}$.
Using the quantum round elimination
lemma (Lemma~\ref{lem:quantroundelim}) again, we get a 
$(2t-2,a+b,a,b)^A$ safe public coin quantum protocol for
\begin{displaymath}
\PAR_{\frac{\log m}{2 c_1 a t^4}, \frac{n}{c_1 b t^4}}
\end{displaymath}
but the error probability increases to at most $\delta + 2 (12t)^{-1}$.

We do the above steps repeatedly. After applying the above steps
$i$ times, we get a 
$(2t-2i,i(a+b),a,b)^A$ safe public coin quantum protocol for
\begin{displaymath}
\PAR_{\frac{\log m}{(2 c_1 a t^4)^i}, \frac{n}{(c_1 b t^4)^i}}
\end{displaymath}
with error probability at most $\delta + 2i (12t)^{-1}$. 

By applying the above steps $t$ times, we finally get a 
$(0,t(a+b),a,b)^A$ safe public coin quantum protocol for
\begin{displaymath}
\PAR_{\frac{\log m}{(2 c_1 a t^4)^t}, \frac{n}{(c_1 b t^4)^t}}
\end{displaymath}
with error probability at most $\delta + 2t (12t)^{-1} < 1/2$.
From the given values of the parameters, we see that
\begin{displaymath}
\frac{\log m}{(2 c_1 a t^4)^t} \geq (\log m)^{\Omega(1)} ~~~~~
\frac{n}{(c_1 b t^4)^t} \geq n^{\Omega(1)}
\end{displaymath}
Thus we get a zero round protocol for a rank parity problem on a
non-trivial domain with error probability less than $1/2$, which
is a contradiction.

In the above proof, we are tacitly ignoring ``rounding off'' problems.
We remark that this does not affect the correctness of the proof.
\end{proof}

\subsection{The `greater-than' problem}
\begin{theorem}
The $t$ round bounded error quantum (classical randomised)
communication complexity of
$GT_n$ is $\Omega(n^{1/t} t^{-3})$ ($\Omega(n^{1/t} t^{-2})$).
\end{theorem}
\begin{proof}
We recall the following reduction from $GT_{n/k}^{(k)}$ to $GT_n$ (see
\cite{miltersen:roundelim}): In $GT_{n/k}^{(k)}$, Alice is given 
$x_1,\ldots,x_k \in \{0,1\}^{n/k}$, Bob is given $i \in [k]$,
$y \in \{0,1\}^{n/k}$, and copies of $x_1,\ldots,x_{i-1}$, and
they have to communicate and decide if $x_i > y$. To reduce 
$GT_{n/k}^{(k)}$ to $GT_n$, Alice constructs 
$\tilde{x} \in \{0,1\}^n$ by concatenating $x_1,\ldots,x_k$, 
Bob constructs $\tilde{y} \in \{0,1\}^n$ by concatenating 
$x_1,\ldots,x_{i-1},y,1^{n(1 - i/k)}$. It is
easy to see that $\tilde{x} > \tilde{y}$ iff $x_i > y$.

Suppose $GT_n$ has a $[t,0,l_1,\ldots,l_t]^A$ safe public coin
quantum protocol with worst case error probability less than $1/3$.
Suppose 
\begin{displaymath}
n \geq \left(C t^3 (l_1 + \cdots + l_t) \right)^t
\end{displaymath}
where $C = (4 \ln 2) 6^4$.
For $1 \leq i \leq t$, define
\begin{displaymath}
k_i \defeq C t^4 l_i ~~~~~
n_i \defeq \frac{n}{\prod_{j=1}^i k_j} ~~~~~
\epsilon_i \defeq \frac{1}{3} + \sum_{j=1}^i 
                  \left(\frac{(4 \ln 2) l_j}{k_j}\right)^{1/4}
\end{displaymath}
Also define $n_0 \defeq n$ and $\epsilon_0 \defeq 1/3$.
Then
\begin{displaymath}
\epsilon_t \defeq \frac{1}{3} + \sum_{j=1}^t 
                  \left(\frac{(4 \ln 2) l_j}{k_j}\right)^{1/4}
              =   \frac{1}{3} + \frac{t}{6t} 
              =   1/2
\end{displaymath}
and
\begin{displaymath}
n_t   =  \frac{n}{\prod_{j=1}^t k_j} 
      =  \frac{n}{(C t^4)^t l_1 \cdots l_t} 
    \geq \frac{n t^t}{C^t t^{4t} (l_1 + \cdots + l_t)^t} 
    \geq 1
\end{displaymath}

We now apply the above self-reduction and
the quantum round elimination
lemma (Lemma~\ref{lem:quantroundelim}) alternately. 
Before the $i$th stage, we have a 
$[t-i+1,\sum_{j=1}^{i-1} l_j,l_i,\ldots,l_t]^Z$ safe public coin
quantum protocol for $GT_{n_{i-1}}$ with
worst case error probability less than $\epsilon_{i-1}$. Here
$Z = A$ if $i$ is odd, $Z = B$ otherwise.
For the $i$th stage, we apply the self-reduction with
$k = k_i$. This gives us a 
$[t-i+1,\sum_{j=1}^{i-1} l_j,l_i,\ldots,l_t]^Z$ safe public coin
quantum protocol for $GT^{(k_i)}_{n_i}$ with the same
error probability. We now apply the quantum round elimination
lemma (Lemma~\ref{lem:quantroundelim}) to get a
$[t-i,\sum_{j=1}^i l_j, l_{i+1},\ldots,l_t]^{Z'}$ safe public coin
quantum protocol for $GT_{n_i}$ with 
worst case error probability less than $\epsilon_i$. Here
$Z' = B$ if $Z = A$ and $Z' = A$ if $Z = B$.
This completes the $i$th stage.

Applying the self-reduction and the round elimination lemma
alternately $t$ times gives us a zero round quantum protocol
for the `greater-than' problem on a domain of size $n_t > 1$
with worst case error probability less than $\epsilon_t = 1/2$, 
which is a contradiction. 

In the above proof, we are tacitly ignoring ``rounding off'' problems.
We remark that this does not affect the correctness of the proof.

This proves the quantum lower bound of $\Omega(n^{1/t} t^{-3})$ on
the message complexity. 

Using the classical round elimination lemma 
(Lemma~\ref{lem:classicalroundelim}) instead of the quantum one,
and treating a classical randomised protocol with complexity $l$
as a $[t,0,l,\ldots,l]^A$ protocol,
we get the stronger classical lower bound of $\Omega(n^{1/t} t^{-2})$.
\end{proof}

Miltersen {\em et al.}~\cite{miltersen:roundelim}
also apply their round elimination lemma to prove (classical)
lower bounds for other data structure problems and communication
complexity problems. We remark that we can extend all those results
in a similar fashion to the quantum world.

\section{Lower bounds for static membership}
\label{sec:sorttable}
Consider the problem of storing a subset $S$ of size at most
$n$ of the universe
$[m]$ in a table with $q$ cells, so that membership queries
can be answered efficiently. We restrict the storage scheme
to be {\em implicit}, using at most $p$ `pointer values'. A
`pointer value' is a member of a set of size $p$ (the set of
`pointers') disjoint from the universe.
The term implicit means that the storage scheme can store
either a `pointer value' or a member of $S$ in a cell.
In particular, the storage scheme
is not allowed to store an element of the universe
which is not a member of $S$.
The query algorithm answers membership queries by performing
$t$ (general) quantum cell probes.
We call such schemes {\em $(p,q,t)$ implicit storage quantum cell
probe schemes}.
For universe sizes $m$ that are `large' compared to $n,p,q$,
we can prove an 
$\Omega(\log n)$ lower bound on the 
number of quantum probes $t$ required to solve
the static membership problem with $(p,q,t)$ implicit storage
quantum cell probe schemes.
We start with the following lemma.
\begin{lemma}
\label{lem:ordsearch}
Suppose $S$ is an $n$ element subset of the universe $[m]$, where 
$m \geq 2n+2$. If the storage scheme is implicit, always stores
the same `pointer' values in the same locations, and in the
remaining locations,
stores the elements of $S$ in a fixed order 
(repetitions of an element are allowed, but all elements have to be
stored) based on their relative ranking in $S$, then
$\Omega(\log n)$ probes are needed by any bounded error
quantum cell query strategy to answer membership queries.
\end{lemma}
\begin{proof} {\bf (Sketch)}
The proof follows by modifying Ambainis's lower bound proof for quantum
ordered searching~\cite{ambainis:binsearch}.
There, it was shown that if $S$ is stored in sorted 
order in a table $T$ then, given any query element $q$, 
$\Omega(\log n)$
probes are required by any quantum search strategy to 
find out the smallest
index $i$, $1 \leq i \leq n$, such that $q \leq T(i)$. We observe that
the lemma above does not follow directly from the result of Ambainis,
since we only need to decide if $q$ is present in the table or not, 
and this is a weaker requirement. To prove the lemma, we follow
the adversary strategy of~\cite{ambainis:binsearch} 
with some minor changes. We study the behaviour of the quantum
query scheme with query element $n+1$.
The proof of Ambainis is based on a clever
strategy of subdividing ``intervals'' (an interval is a contiguous
set of locations in the sorted table). We work instead with 
``logical intervals'', where a logical interval denotes the set of
locations in the table where elements contiguous in the natural
ordering are stored (as determined  by the fixed storing order).
After this definition, one can easily show that
the same subdivision strategy as in~\cite{ambainis:binsearch}
goes through.
In Ambainis's proof, the adversary constructs inputs by padding with 
zeros from the beginning up to the left of 
an interval, and with ones from the end up to the right of the interval.
Instead, we pad 
with {\em small} numbers ($1,2, \ldots$) from the logical 
beginning up to the logical
left of a logical interval, and with {\em large} numbers 
($m,m-1, \ldots$) from the 
logical end up to the  logical right of the logical 
interval. We store the appropriate `pointer values' in the `pointer
locations'
(predetermined by the storing strategy). After doing this, one can
easily show that the same error 
analysis of~\cite{ambainis:binsearch} goes through. 
Thus, the adversary
finally can produce two inputs, one of them containing $n+1$ 
and the other not, such that the behaviour of the query scheme is very
similar on both. This is a contradiction. 
\end{proof}

\paragraph{Remark:} H{\o}yer {\em et al.} also prove an
$\Omega(\log n)$ lower bound
for quantum ordered searching~\cite{hoyer:binsearch}. But their
approach, which is based on ``distinguishing
oracles'', does not seem to be suitable for 
proving lower bounds for boolean valued functions. Hence to prove
Lemma~\ref{lem:ordsearch}, we modify
the older $\Omega(\log n)$ lower bound of Ambainis for quantum
ordered searching.

\begin{theorem}
\label{thm:sorttable}
For every $n,p,q$, there exists an $N(n,p,q)$ such that for all 
$m \geq N(n,p,q)$, the following holds: Consider any bounded error
$(p,q,t)$ implicit storage quantum cell probe scheme for the 
static membership problem with universe size $m$ and size of
the stored subset at most $n$. Then the quantum query scheme must make
$t = \Omega(\log n)$ probes.
\end{theorem}
\begin{proof} {\bf (Sketch)}
Our proof follows from the Ramsey theoretic arguments of 
Yao~\cite{yao:tablesort} together with Lemma~\ref{lem:ordsearch}.
The details are omitted. 
\end{proof}

\section{Conclusions and open problems}
\label{sec:conclusion}
In this paper 
we introduce the quantum cell probe model, a model for 
studying static data structure problems in the quantum world. 
We show that
the additional power of quantum querying does not help for 
the static membership problem when the storage scheme 
is restricted to be implicit, generalising a result of Yao. 
We also explore the possibility of using quantum 
communication complexity
to prove lower bounds in the quantum cell probe model.
We prove a round elimination lemma for quantum communication complexity 
and use it to prove lower bounds for the static predecessor problem in 
a restricted version of the quantum cell probe model, the address-only 
version. Extending this result to the general model
remains an important
open problem. We also use the quantum round elimination lemma to prove
rounds versus communication tradeoffs for the `greater-than'
problem. It would be interesting to find
other applications of the 
round elimination lemma to quantum communication complexity.

\subsection*{Acknowledgements} We thank Ashwin Nayak, 
Jaikumar Radhakrishnan, Rahul Jain, Hartmut Klauck  and 
Peter Bro Miltersen for helpful discussions and feedback. 
We also thank Jaikumar
Radhakrishnan for reading an early draft of this paper 
and helping us to improve the presentation of this paper.

\bibliography{quantcell}

\appendix
\section*{Appendix}

\section{Proof of Lemma~\ref{lem:classicalroundreduce}}
In this section, we prove Lemma~\ref{lem:classicalroundreduce}. 
The proof is somewhat similar to the proof of Lemma~4.4 in Klauck 
{\em et al.}~\cite{klauck:pointer}, but much simpler since we
are in the classical setting.
We first state a theorem which will be required in the proof of 
Lemma~\ref{lem:classicalroundreduce}. The quantum version of this
theorem, called the ``average encoding theorem'', has been 
proved by Klauck {\em et al.}~\cite{klauck:pointer}, who also
use it in the proof of Lemma~4.4 in their paper.
Intuitively speaking, the theorem says that if the mutual information 
between
a (classical) random variable and its (classical) encoding is small, 
then the various probability distributions on the codewords 
are close to the average probability distribution on the codewords.
Below, the notation $\ellone{\sigma - \rho}$ stands 
for the total variation
distance ($\ell_1$ distance) between probability distributions
$\sigma$ and $\rho$ over the same sample space.
\begin{theorem}[Average encoding, classical version, 
                \cite{klauck:pointer}]
\label{thm:averageencodingclassical}
Let $X$ be a classical
random variable, which takes value $x$ with probability $p_x$, and
$M$ be a classical randomised encoding $x \mapsto \sigma_x$ of $X$,
where $\sigma_x$ is a probability distribution
over the sample space of codewords. The probability distribution of the
average encoding is $\sigma \defeq \sum_x p_x \sigma_x$. 
Then
\begin{displaymath}
\sum_x p_x \ellone{\sigma_x - \sigma} \leq \sqrt{(2 \ln 2) I(X:M)}
\end{displaymath}
\end{theorem}

We now proceed to the proof of Lemma~\ref{lem:classicalroundreduce}. 

\ \\
{\bf Lemma~\ref{lem:classicalroundreduce}}
{\it
Suppose $f: E \times F \rightarrow G$ is a function. Let $D$ be
a probability distribution on $E \times F$, and
$P$ be a $[t,0,l_1,\ldots,l_t]^A$ private coin
classical randomised protocol for $f$. 
Let $X$ stand for the classical random variable denoting Alice's input
(under distribution $D$),
$M$ be the first message of Alice in the protocol $P$, and 
$I(X:M)$ denote the mutual information between $X$ and $M$ under
distribution $D$. Then
there exists a $[t-1,0,l_2,\ldots,l_t]^B$ public coin
classical randomised protocol $Q$ for $f$, such that
\begin{displaymath}
\epsilon^Q_D \leq \epsilon^P_D + \frac{1}{2} ((2 \ln 2) I(X:M))^{1/2}
\end{displaymath}
\/} \\
\begin{proof}
We first give an overview of the plan of the proof, before
getting down to the details. The proof proceeds in stages.

\paragraph{Stage 1:} Starting from $P$, we construct a 
$[t,l_1,\ldots,l_t]^A$ private coin protocol $P'$, 
where the first message
is independent of Alice's input, and $\epsilon^{P'}_D 
\leq \epsilon^{P}_D + (1/2) ((2 \ln 2) I(X:M))^{1/2}$.
The important idea in this step is to first generate Alice's
message using a new private coin without ``looking'' at her input, 
and after that,
to adjust Alice's old private coin in a suitable manner so as to
be consistent with her message and input.

\paragraph{Stage 2:} Suppose the coin tosses in $P'$ were done in
public. Then Bob can generate the first message of $P'$
himself, as it is independent of Alice's input. 
Doing this gives us a $[t-1,l_2,\ldots,l_t]^B$ public coin
protocol $Q$, such that
$\epsilon^Q_{x,y} = \epsilon^{Q'}_{x,y}$ for
every $(x,y) \in E \times F$. 

The protocol $Q$ of Stage~2
is our desired $[t-1,l_2,\ldots,l_t]^B$ public coin classical 
randomised protocol for $f$. We have 
\begin{displaymath}
\epsilon^Q_D = \epsilon^{P'}_D 
   \leq \epsilon^P_D + \frac{1}{2} ((2 \ln 2) I(X:M))^{1/2} 
\end{displaymath}

We now give the details of the proof.
Let $\sigma_x$ be the probability distribution of the first message 
$M$ of protocol $P$ when Alice's input $X=x$. 
Let $Y$ denote Bob's input register.
Define $\sigma \defeq \sum_x p_x \sigma_x$, where
$p_x$ is the (marginal) probability of $x$ under distribution $D$.
$\sigma$ is the probability
distribution of the average first message under distribution $D$. 
By Theorem~\ref{thm:averageencodingclassical}, we get that
\begin{displaymath}
\sum_x p_x \ellone{\sigma_x - \sigma} \leq \sqrt{(2 \ln 2) I(X:M)}
\end{displaymath}

For $x \in E$ and an instance $m$ of the first message of Alice, let 
$q^{xm}_r$ denote the (conditional) probability that the private
coin toss of Alice results in $r$, given that Alice's input is $x$
and her first message in protocol $P$ is $m$. Let $\sigma(m \mid x)$ 
denote the probability that the first message of Alice in $P$ is $m$,
given that her input is $x$. Let $\sigma(m)$ denote the 
probability of $m$ occurring in the average first message of Alice.
Then, $\sigma(m) = \sum_x p_x \sigma(m \mid x)$.

\paragraph{Stage 1:}
We construct a $[t,0,l_1,\ldots,l_t]^A$ private coin classical
randomised protocol $P'$ for $f$ with average error under 
distribution $D$,
$\epsilon^{P'}_D \leq \epsilon^P_D + (1/2) ((2 \ln 2) I(X:M))^{1/2}$,
and where the probability distribution of
the first message is independent of the input to Alice.
Suppose Alice is given
$x \in E$ and Bob is given $y \in F$. Alice tosses a fresh private
coin to pick $m$ with probability $\sigma(m)$. She then sets
her old private coin to $r$ with probability $q^{xm}_r$. (If in
$P$, message $m$ cannot occur when Alice's input is $x$, we
say that protocol $P'$ gives an error if such a thing happens.)
After this, Alice and Bob behave as in protocol $P$ (henceforth,
Alice ignores the new private coin which she had tossed to generate
her first message $m$). 
Hence in $P'$, the probability distribution of the first message
is independent of Alice's input. 

Let us now compare the situations in protocols
$P$ and $P'$ when Alice's input is $x$, Bob's input is $y$, 
Alice has finished tossing her private coins, but no communication
has taken place as yet. 
In protocol $P$, the probability that Alice's private coin toss 
results in $r$ is 
\begin{displaymath}
\sum_m \sigma(m \mid x) q^{xm}_r
\end{displaymath}
In protocol $P'$, the probability that Alice's (old) private coin toss 
results in $r$ is 
\begin{displaymath}
\sum_m \sigma(m) q^{xm}_r
\end{displaymath}
Thus, the $\ell_1$ distance between the probability distributions
on Alice's (old) private coin toss is
\begin{eqnarray*}
\lefteqn{\sum_r \left|\sum_m q^{xm}_r (\sigma(m \mid x)-\sigma(m))
                \right|} \\
     & \leq & \sum_r \sum_m q^{xm}_r \left| \sigma(m \mid x) - 
                                            \sigma(m) 
                                     \right| \\
     &   =  & \sum_m \left(\left| \sigma(m \mid x) - \sigma(m) \right| 
                            \sum_r q^{xm}_r  \right) \\
     &   =  & \sum_m \left| \sigma(m \mid x) - \sigma(m) \right| \\
     &   =  & \ellone{\sigma_x - \sigma} 
\end{eqnarray*}
Hence, the error probability of $P'$ on input $x,y$ 
\begin{displaymath}
\epsilon^{P'}_{x,y} \leq \epsilon^P_{x,y} + 
                         \frac{1}{2} \ellone{\sigma_x - \sigma}
\end{displaymath}
Let $q_{xy}$ be the probability that $(X,Y)=(x,y)$ under distribution
$D$. Then, the average error of $P'$ under distribution $D$,
$\epsilon^{P'}_D$, is bounded by
\begin{eqnarray*}
\epsilon^{P'}_D 
  &   =  & \sum_{x,y} q_{xy} \epsilon^{P'}_{x,y} \\
  & \leq & \sum_{x,y} q_{xy} \left(\epsilon^P_{x,y} + \frac{1}{2} 
                                   \ellone{\sigma_x - \sigma}\right) \\
  &   =  & \epsilon^P_D + 
           \frac{1}{2} \sum_x p_x \ellone{\sigma_x - \sigma}   \\
  & \leq & \epsilon^P_D + \frac{1}{2} ((2 \ln 2) I(X:M))^{1/2} 
\end{eqnarray*}
The last inequality follows from the ``average
encoding theorem'' (Theorem~\ref{thm:averageencodingclassical}).

\paragraph{Stage 2:}
We now construct our desired $[t-1,0,l_2,\ldots,l_t]^B$ 
public coin classical randomised protocol $Q$ for $f$ with
$\epsilon^Q_D = \epsilon^{P'}_D$. Suppose all the coin tosses 
of Alice and Bob in $P'$ were done publicly before any communication
takes place. Now there is no need for the first message from 
Alice to Bob, because Bob can reconstruct the message by looking
at the public coin tosses. This gives us the protocol $Q$, and
trivially
\begin{displaymath}
\epsilon^Q_D = \epsilon^{P'}_D 
   \leq  \epsilon^P_D + \frac{1}{2} ((2 \ln 2) I(X:M))^{1/2} 
\end{displaymath}

This completes the proof of Lemma~\ref{lem:classicalroundreduce}.
\end{proof}

\section{Proof of Lemma~\ref{lem:quantroundreduce}}
In this section, we prove Lemma~\ref{lem:quantroundreduce}. 
We first start with the definition of the trace norm of linear
operators, then state
three theorems which will be required in the proof of 
Lemma~\ref{lem:quantroundreduce}, and after that, we finally 
present the proof of Lemma~\ref{lem:quantroundreduce}.

For a linear operator $A$ on a finite dimensional Hilbert space,
the {\em trace norm} of $A$ is defined as 
$\trace{A} \defeq \Tr \sqrt{A^\dagger A}$.
The following fundamental theorem (see~\cite{aharonov:mixed})
shows that the
trace distance between two density matrices $\rho_1,\rho_2$, 
$\trace{\rho_1 - \rho_2}$, bounds how well one can distinguish
between $\rho_1,\rho_2$ by a measurement.
\begin{theorem}[\cite{aharonov:mixed}]
\label{thm:l1tracedist}
Let $\rho_1,\rho_2$ be two density matrices on the same Hilbert
space. Let ${\cal M}$ be a general measurement (i.e. a POVM),
and ${\cal M} \rho_i$ denote the probability distributions on
the (classical) outcomes of ${\cal M}$ got by performing measurement
${\cal M}$ on $\rho_i$. Let the
$\ell_1$ distance between ${\cal M} \rho_1$ and ${\cal M} \rho_2$ 
be denoted by
$\ellone{ {\cal M} \rho_1 - {\cal M} \rho_2 }$. Then
\begin{displaymath}
\ellone{ {\cal M} \rho_1 - {\cal M} \rho_2 } \leq 
                                     \trace{\rho_1 - \rho_2}
\end{displaymath}
\end{theorem}

In the proof of Lemma~\ref{lem:quantroundreduce},
we will need the following ``average encoding theorem'' of
Klauck {\em et al.}~\cite{klauck:pointer}. 
Intuitively speaking, it says that if the mutual information between
a classical random variable and its quantum encoding is small, then
the various quantum ``codewords'' are close to the ``average codeword''.
\begin{theorem}[Average encoding, quantum version, 
                \cite{klauck:pointer}]
\label{thm:averageencodingquantum}
Suppose $X$, $Q$ are two disjoint quantum systems, 
where $X$ is a classical
random variable, which takes value $x$ with probability $p_x$, and
$Q$ is a quantum encoding $x \mapsto \sigma_x$ of $X$. Let the density
matrix of the average encoding be $\sigma \defeq \sum_x p_x \sigma_x$. 
Then
\begin{displaymath}
\sum_x p_x \trace{\sigma_x - \sigma} \leq \sqrt{(2 \ln 2) I(X:Q)}
\end{displaymath}
\end{theorem}

We will also need the following ``local transition theorem'' of
Klauck {\em et al.}~\cite{klauck:pointer}.
\begin{theorem}[Local transition, \cite{klauck:pointer}]
\label{thm:localtransition}
Let $\rho_1,\rho_2$ be two mixed states with support in a Hilbert
space ${\cal H}$, ${\cal K}$ any Hilbert space of dimension at least
the dimension of ${\cal H}$, and $\ket{\phi_i}$ any purifications
of $\rho_i$ in ${\cal H} \otimes {\cal K}$. Then, there is a local
unitary transformation $U$ on ${\cal K}$ that maps $\ket{\phi_2}$
to $\ket{\phi_2'} \defeq (I \otimes U) \ket{\phi_2}$ ($I$ is the
identity operator on ${\cal H}$) such that
\begin{displaymath}
\trace{\ketbra{\phi_1} - \ketbra{\phi_2'}} 
        \leq 2 \sqrt{\trace{\rho_1 - \rho_2}}
\end{displaymath}
\end{theorem}

We now proceed to the proof of Lemma~\ref{lem:quantroundreduce}. The
proof is similar to the proof of Lemma~4.4 in
\cite{klauck:pointer}, but with a careful accounting of 
``safe'' overheads in the messages communicated by Alice and Bob. 

\ \\
{\bf Lemma~\ref{lem:quantroundreduce}}
{\it
Suppose $f: E \times F \rightarrow G$ is a function. Let $D$ be
a probability distribution on $E \times F$, and
$P$ be a $[t,c,l_1,\ldots,l_t]^A$ safe coinless 
quantum protocol for $f$. 
Let $X$ stand for the classical random variable denoting Alice's input
(under distribution $D$),
$M$ be the first message of Alice in the protocol $P$, and 
$I(X:M)$ denote the mutual information between $X$ and $M$ under
distribution $D$. Then
there exists a $[t-1,c+l_1,l_2,\ldots,l_t]^B$ safe coinless 
quantum protocol $Q$ for $f$, such that
\begin{displaymath}
\epsilon^Q_D \leq \epsilon^P_D + ((2 \ln 2) I(X:M))^{1/4}
\end{displaymath} 
\/} \\
\begin{proof}
We first give an overview of the plan of the proof, before
getting down to the details. The proof proceeds in stages.
We remark on the similarities between the stages in the
quantum proof, and the stages in the classical proof
(Lemma~\ref{lem:classicalroundreduce}). 
Stages~1A and 1B of the quantum proof together correspond to
Stage~1 of the classical proof, and
Stages~2A and 2B of the quantum proof together correspond to
Stage~2 of the classical proof.

\paragraph{Stage 1A:}
Starting from the $[t,c,l_1,\ldots,l_t]^A$ safe 
coinless protocol $P$, we construct a $[t,c,l_1,\ldots,l_t]^A$ 
safe coinless protocol $\tilde{P}$ with 
$\epsilon^{\tilde{P}}_{x,y} = \epsilon^P_{x,y}$ for
every $(x,y) \in E \times F$. $\tilde{P}$
contains an extra ``secure'' copy of Alice's input
$x \in E$, but is otherwise the same as $P$. 

\paragraph{Stage 1B:}
Starting from $\tilde{P}$, we construct a 
$[t,c,l_1,\ldots,l_t]^A$ safe coinless protocol $P'$, 
where the first message
is independent of Alice's input, and $\epsilon^{P'}_D 
\leq \epsilon^{\tilde{P}}_D + ((2 \ln 2) I(X:M))^{1/4}$.
The important idea in this step is to first generate Alice's
average message (which is independent of her input), 
and after that, use the extra ``secure'' copy of 
Alice's input $x$ to apply
a unitary transformation $U_x$ on some of her qubits without
touching her message.
$U_x$ is used to adjust Alice's state in a suitable
manner so as to be consistent with her input and message.
This ``adjustment'' step requires the use of the ``local transition 
theorem'' (Theorem~\ref{thm:localtransition}).

\paragraph{Stage 2A:}
Since in $P'$ the first message is independent of
Alice's input, Bob can generate it himself. But it
is also necessary
to achieve the correct entanglement between
Alice's qubits and the first message. Bob does this
by first sending a safe message of $l_1+c$ qubits.
Alice then applies a unitary transformation 
$V_x$ on some of her qubits, using the extra ``secure''
copy of her input $x$, to achieve the correct entanglement. 
The existence of such a $V_x$ follows from
Theorem~\ref{thm:localtransition}.
Doing all this gives us a 
$[t+1,c+l_1,0,0,l_2,\ldots,l_t]^B$ safe coinless 
protocol $Q'$, such that  
$\epsilon^{Q'}_{x,y} = \epsilon^{P'}_{x,y}$ for
every $(x,y) \in E \times F$. 

\paragraph{Stage 2B:}
Since the first message of Alice in $Q'$ is zero qubits
long, Bob can concatenate his first two messages, giving
us a $[t-1,c+l_1,l_2,\ldots,l_t]^B$ safe coinless
protocol $Q$, such that
$\epsilon^Q_{x,y} = \epsilon^{Q'}_{x,y}$ for
every $(x,y) \in E \times F$. The technical reason
behind this is that unitary transformations on disjoint
sets of qubits commute. 

The protocol $Q$ of Stage~2B
is our desired $[t-1,c+l_1,l_2,\ldots,l_t]^B$ safe coinless quantum 
protocol for $f$. We have 
\begin{displaymath}
\epsilon^Q_D = \epsilon^{Q'}_D = \epsilon^{P'}_D 
   \leq  \epsilon^{\tilde{P}}_D + ((2 \ln 2) I(X:M))^{1/4} 
     =   \epsilon^P_D + ((2 \ln 2) I(X:M))^{1/4} 
\end{displaymath}

We now give the details of the proof.
Let $\sigma_x$ be the density matrix of the first message $M$ of
protocol $P$ when Alice's input $X=x$. Let $Y$ denote Bob's input
register.
Define $\sigma \defeq \sum_x p_x \sigma_x$, where
$p_x$ is the (marginal) probability of $x$ under distribution $D$.
$\sigma$ is the density
matrix of the average first message under distribution $D$. By
the ``secureness'' of $P$, $\sigma$ is also
the density matrix of the first message when $\ket{\psi}$ is 
fed to Alice's input register $X$, where 
$\ket{\psi} \defeq \sum_x \sqrt{p_x} \ket{x}$.
By Theorem~\ref{thm:averageencodingquantum}, we get that
\begin{displaymath}
\sum_x p_x \trace{\sigma_x - \sigma} \leq \sqrt{(2 \ln 2) I(X:M)}
\end{displaymath}

\paragraph{Stage 1A:} 
We first construct a $[t,c,l_1,\ldots,l_t]^A$ safe coinless quantum
protocol $\tilde{P}$ for $f$ such that
$\epsilon^{\tilde{P}}_{x,y} = \epsilon^P_{x,y}$, for
every $(x,y) \in E \times F$. 
Let $X$ be Alice's input register in $P$. 
In $\tilde{P}$, Alice has an additional register $C$, and the input
$x$ to Alice is fed to register $C$, instead of $X$. $X$ is initialised
to $\ket{0}$ in $\tilde{P}$. In protocol $\tilde{P}$, Alice first
copies the contents of $C$ to $X$. After that, things in $\tilde{P}$
proceed as in $P$. Register $C$ is not touched henceforth, and thus,
$C$ holds an extra ``secure'' copy of $x$ throughout the run of
protocol $\tilde{P}$.

\paragraph{Stage 1B:} 
We now construct a $[t,c,l_1,\ldots,l_t]^A$ safe coinless quantum
protocol $P'$ for $f$ with average error under distribution $D$,
$\epsilon^{P'}_D \leq \epsilon^{\tilde{P}}_D + 
                      ((2 \ln 2) I(X:M))^{1/4}$,
and where the density matrix of
the first message is independent of the input $x$ to Alice.
Alice is given
$x \in E$ and Bob is given $y \in F$.
Consider the situation in $\tilde{P}$
after the first message has been prepared by Alice,
but before it is sent to Bob. Let register $A$ denote Alice's qubits
excluding the message qubits $M$ and the qubits of the
``secure'' copy $C$ (in particular, $A$ includes the qubits of
register $X$).
Without loss of generality, one can assume that register $A$ has
at least $l_1+c$ qubits, because one can initially pad up $A$ with
ancilla qubits set to $\ket{0}$.
Let $\ket{x}_C \otimes \ket{\theta_x}_{AM}$ be the state vector of
$CAM$ in $\tilde{P}$ at this point,
where the subscripts denote the registers.
$\ket{\theta_x}_{AM}$ is a purification of $\sigma_x$.
We note that $\ket{\theta_x}$ is also the state vector of
$AM$ in protocol $P$ at this point.
$P'$ is similar to $\tilde{P}$ except for the following. 
Alice puts $\ket{\psi}$ in register $X$ (instead of copying
$C$ to $X$ as in $\tilde{P}$) to create
the first message in register $M$ with
density matrix $\sigma$. $AM$ now contains a purification 
$\ket{\theta}$ of
$\sigma$. Then Alice applies a unitary transformation
$U_x$ depending upon $x$ (which is available ``securely''
in register $C$) on $A$,
so that $\ket{\theta_x'}_{AM} \defeq (U_x \otimes I) \ket{\theta}_{AM}$ 
is ``close'' to $\ket{\theta_x}_{AM}$. Here $I$ stands for
the identity transformation on $M$.
Theorem~\ref{thm:localtransition} tells
us that there exists a unitary transformation $U_x$ on $A$ 
such that
\begin{displaymath}
\trace{\ketbra{\theta_x} - \ketbra{\theta_x'}} 
        \leq 2 \sqrt{\trace{\sigma_x - \sigma}}
\end{displaymath}
Thus, $\ket{x}_C \otimes \ket{\theta_x'}_{AM}$ is the state vector of 
$CAM$ in $P'$ after the application of $U_x$. 
Alice then sends register $M$ to Bob and
after this, Alice and Bob behave as in $\tilde{P}$.
Application of $U_x$ does not affect the
density matrix of register $M$, which continues to
be $\sigma$. Hence in $P'$, the density matrix of the first message
is independent of Alice's input. 

Let us now compare the situations in protocols
$\tilde{P}$ and $P'$ when Alice's input is $x$, Bob's input is $y$, 
Alice has prepared her first message, but no communication
has taken place as yet. 
At this point, in both protocols $\tilde{P}$ and $P'$,
the state vector of Bob's qubits is the same, and in
tensor with the state vector of Alice's qubits. Let $B$ denote the
register of Bob's qubits (including his input qubits $Y$)
and let $\ket{\eta}_B$ denote
the state vector of $B$ at this point. Hence the global
state of protocol $\tilde{P}$ at this point is 
$\ket{x}_C \otimes \ket{\theta_x}_{AM}  \otimes \ket{\eta}_B$, and 
the global state of $P'$ is 
$\ket{x}_C \otimes \ket{\theta_x'}_{AM} \otimes \ket{\eta}_B$.
Therefore, the global
states of protocols $\tilde{P}$ and $P'$ at this point differ
in trace distance by the quantity
\begin{displaymath}
\trace{\ketbra{x} \otimes \ketbra{\theta_x}  
                  \otimes \ketbra{\eta} - 
       \ketbra{x} \otimes \ketbra{\theta_x'} 
                  \otimes \ketbra{\eta}}
= \trace{\ketbra{\theta_x} - \ketbra{\theta_x'}} 
        \leq 2 \sqrt{\trace{\sigma_x - \sigma}}
\end{displaymath}
Using Theorem~\ref{thm:l1tracedist}, we see that
the error probability of $P'$ on input $x,y$ 
\begin{displaymath}
\epsilon^{P'}_{x,y} \leq \epsilon^{\tilde{P}}_{x,y} + 
               \frac{1}{2} \trace{\ketbra{x} \otimes \ketbra{\theta_x}
                                      \otimes \ketbra{\eta} -
                                 \ketbra{x} \otimes \ketbra{\theta_x'}
                                      \otimes \ketbra{\eta}}
                    \leq \epsilon^{\tilde{P}}_{x,y} + 
                           \sqrt{\trace{\sigma_x - \sigma}}
\end{displaymath}
Let $q_{xy}$ be the probability that $(X,Y)=(x,y)$ under distribution
$D$. Then, the average error of $P'$ under distribution $D$,
$\epsilon^{P'}_D$, is bounded by
\begin{eqnarray*}
\epsilon^{P'}_D 
  &   =  & \sum_{x,y} q_{xy} \epsilon^{P'}_{x,y} \\
  & \leq & \sum_{x,y} q_{xy} \left(\epsilon^{\tilde{P}}_{x,y} +
                                   \sqrt{\trace{\sigma_x - \sigma}}
                             \right)  \\
  & \leq & \epsilon^{\tilde{P}}_D + \sqrt{\sum_{x,y} q_{xy}
			                  \trace{\sigma_x - \sigma}} \\
  &  =   & \epsilon^{\tilde{P}}_D + \sqrt{\sum_x p_x
			                  \trace{\sigma_x - \sigma}} \\
  & \leq & \epsilon^{\tilde{P}}_D + ((2 \ln 2) I(X:M))^{1/4} 
\end{eqnarray*}
For the second inequality above, we use the concavity of the square
root function. The last inequality follows from the ``average
encoding theorem'' (Theorem~\ref{thm:averageencodingquantum}).

\paragraph{Stage 2A:} 
We now construct a $[t+1,c+l_1,0,0,l_2,\ldots,l_t]^B$ safe 
coinless quantum protocol $Q'$ for $f$ with
$\epsilon^{Q'}_{x,y} = \epsilon^{P'}_{x,y}$, for all
$(x,y) \in E \times F$. 
Alice is given $x \in E$ and Bob is given $y \in F$. 
The protocol $Q'$ will be constructed from $P'$.
The input $x$ is fed to register $C$ of Alice, and the input
$y$ is fed to register $Y$ of Bob. Let register $G$ denote
all the qubits of register $A$, except the last $l_1+c$ qubits.
In protocol $Q'$ the registers initially in Alice's possession 
are $C$ and $G$, and the registers initially in Bob's possession
are $B$, $M$, and a new register $R$, where $R$ is
$l_1+c$ qubits long. 
The qubits of $G$ are initially set to $\ket{0}$.
Bob first prepares the state vector $\ket{\eta}$ in register $B$ as
in protocol $P'$. He then
constructs a canonical purification of $\sigma$ 
in registers $MR$. The density matrix of $M$ is $\sigma$.
Bob then sends $R$ to Alice. The density matrix of $R$ is independent 
of the inputs $x,y$ (in fact, if the canonical purification
in $MR$ is the Schmidt purification, then the density matrix of
$R$ is also $\sigma$). 
After receiving $R$, Alice treats $GR$ as the register $A$ in the
remainder of the protocol. $A M$ now contains a purification of
$\sigma$. Alice applies a unitary transformation
$V_x$ depending upon $x$ (which is available ``securely'' in register
$C$) on $A$,
so that the state vector of $AM$ becomes
$\ket{\theta_x'}_{AM}$. The existence of such
a $V_x$ follows from Theorem~\ref{thm:localtransition}.
At this point, the global state vector
(over all the qubits of Alice and Bob) in $Q'$ is the same as the
global state vector in $P'$ viz. 
$\ket{x}_C \otimes \ket{\theta_x'}_{AM} \otimes \ket{\eta}_B$. 
Bob now treats register $M$ as if
it were the first message of Alice in $P'$, and proceeds to compute
his response $N$ of length $l_2$. Bob sends $N$ to Alice and after this 
protocol $Q'$ proceeds as in $P'$. In $Q'$
Bob starts the communication, the communication
goes on for $t+1$ rounds, the first message of Bob of length
$l_1+c$ (i.e. register $R$) is a safe message, and the first
message of Alice is zero qubits long. 

\paragraph{Stage 2B:} 
We finally construct a $[t-1,c+l_1,l_2,\ldots,l_t]^B$ safe 
coinless quantum protocol $Q$ for $f$ with
$\epsilon^Q_{x,y} = \epsilon^{Q'}_{x,y}$, for all
$(x,y) \in E \times F$. 
In protocol $Q$, Bob (after doing the same computations as in $Q'$)
first sends as a single
message register $RN$ of length $(l_1+c)+l_2$,
and after that
Alice applies $V_x$ on $A$ followed by her appropriate
unitary transformation 
on $AN$ (the unitary transformation of Alice in $Q'$ on
her qubits $AN$ after she has received the first two messages of Bob).
At this point, the global state vector (over all the qubits of
Alice and Bob) in $Q$ is the same as the global state vector in $Q'$,
since unitary transformations on disjoint sets of qubits commute.
After this, things in $Q$ proceed as in $Q'$. 
In protocol $Q$ Bob starts the communication, the 
communication
goes on for $t-1$ rounds, and the first message of Bob of length
$(l_1+c)+l_2$ contains a safe overhead (the register $R$) of 
$l_1+c$ qubits. 

This completes the proof of Lemma~\ref{lem:quantroundreduce}.
\end{proof}

\end{document}